\documentclass{ametsocV6.1}
\usepackage{subcaption}
\usepackage[hidelinks]{hyperref}
\usepackage[capitalize]{cleveref}
\usepackage{natbib}
\usepackage{booktabs}
\usepackage{multirow}
\usepackage{float}
\usepackage{amsmath}

\usepackage{amssymb}
\usepackage{setspace}
\title{ArchesClimate: Probabilistic Decadal Ensemble Generation With Flow Matching}

\authors{Graham Clyne\aff{a}\correspondingauthor{Graham Clyne, clyne.graham@gmail.com}, Guillaume Couairon\aff{a}, Guillaume Gastineau\aff{b}, Claire Monteleoni\aff{a,c}, Anastase Charantonis\aff{a}}

\affiliation{\aff{a}{ARCHES, INRIA, Paris, France},\aff{b}{UMR LOCEAN, IPSL, Sorbonne Université, IRD, CNRS, MNHN, Paris, France},\aff{c}{Department of Computer Science, University of Colorado Boulder, USA}}

\abstract{Internal variability is a dominant contributor to the uncertainty of predictions at the interannual to decadal timescale. A typical approach to separating the internal variability from forced climate responses is to generate large ensembles of simulations under different initial conditions. Due to the complexity of Earth System Models, generating these large ensembles is computationally expensive. In this work, we present ArchesClimate, a deep learning-based climate model emulator designed to reduce the cost of exploring internal variability at timescales ranging from monthly to decadal. ArchesClimate is trained on decadal hindcasts of the IPSL-CM6A-LR climate model. We train a flow matching model following ArchesWeatherGen \citep{couaironArchesWeatherGenSkillfulComputeefficient2026}, which we adapt to predict near-term climate. Once trained, the model generates states at a one-month lead time from the states of the two preceding months, and can be used to auto-regressively emulate climate model simulations. We show that for up to 10 years, these generations are stable and physically consistent. We also show that for several important climate variables, ArchesClimate generates simulations that are interchangeable with the IPSL model. This work suggests that climate model emulators could reduce the cost of generating large ensembles with climate models. }

\begin{document}

%% Necessary!
\maketitle

%%%%%%%%%%%%%%%%%%%%%%%%%%%%%%%%%%%%%%%%%%%%%%%%%%%%%%%%%%%%%%%%%%%%%
% SIGNIFICANCE STATEMENT/CAPSULE SUMMARY
%%%%%%%%%%%%%%%%%%%%%%%%%%%%%%%%%%%%%%%%%%%%%%%%%%%%%%%%%%%%%%%%%%%%%
%
% If you are including an optional significance statement for a journal article or a required capsule summary for BAMS 
% (see www.ametsoc.org/ams/index.cfm/publications/authors/journal-and-bams-authors/formatting-and-manuscript-components for details), 
% please apply the necessary command as shown below:
%
% Significance Statement (all journals except BAMS)
%
%\statement
%	 Enter significance statement here, no more than 120 words. See \url{www.ametsoc.org/index.cfm/ams/publications/author-information/significance-statements/} for details.
%
%% Capsule (BAMS only)
%%
%\capsule
%       Enter BAMS capsule here, no more than 30 words. See \url{www.ametsoc.org/index.cfm/ams/publications/author-information/formatting-and-manuscript-components/#capsule} for details.
%
%% * * If using twocol mode, you will need to use the commands "twocolsig" and "twocolcapsule" in place of "sig" and "capsule"
%%      to ensure that the text box correctly spans across both columns.
%

%%%%%%%%%%%%%%%%%%%%%%%%%%%%%%%%%%%%%%%%%%%%%%%%%%%%%%%%%%%%%%%%%%%%%
% MAIN BODY OF PAPER
%%%%%%%%%%%%%%%%%%%%%%%%%%%%%%%%%%%%%%%%%%%%%%%%%%%%%%%%%%%%%%%%%%%%%
%

\section{Introduction}

Ensemble generation is an important tool to investigate climate. Given the chaotic nature of atmospheric and oceanic dynamics, changes in initial conditions can lead to distinct climates. Ensemble simulation can be seen as repeating an experimentation protocol with the same boundary conditions using different initial conditions that enable the exploration of internal variability. Internal variability in climate models refers to the natural fluctuations in the climate system that occur without variations from external forcings. By repeatedly sampling the simulated climate, we enable probabilistic analyses that support the investigation of a broad range of problems, including extreme event attribution and uncertainty quantification \citep{fyfeLargeNeartermProjected2017,fischerStorylinesUnprecedentedHeatwaves2023}. 

Furthermore, generating large ensembles (e.g. 30-100 members) can be used to separate the internal variability from the climate response to external forcings such as greenhouse gases or solar irradiation \citep{maherLargeEnsembleClimate2021,eadeSeasonaltodecadalClimatePredictions2014, deserInsightsEarthSystem2020}.  These fluctuations arise from complex interactions between the atmosphere, ocean, land, and ice (e.g. El Niño or the North Atlantic Oscillation) and can cause variations in climate at all time scales. While external forcings, scenario choices, and model structure uncertainties dominate climate projections at multi-decadal to centennial timescales, internal variability remains the dominant source of uncertainty at decadal timescales (up to 10 years) and at regional scales \citep{hawkinsPotentialNarrowUncertainty2009, lehnerPartitioningClimateProjection2020a}. 

Generating ensembles requires running multiple instances of climate models \citep{deserEffectsMacroVs2024}. This practice is computationally expensive and severely limits the generation of large ensembles with more than 100 members \citep{maherLargeEnsembleClimate2021}. To overcome this hurdle, significant research has focused on developing efficient climate model emulators that can reduce the associated computational cost.

Significant progress has been made in probabilistic climate emulation, notably with models like NeuralGCM \citep{kochkovNeuralGeneralCirculation2024}. This model combines an atmospheric dynamical core and a data-driven atmospheric physical core. Through the injection of noise into the encoder and the data-driven core, NeuralGCM was trained to reproduce an ensemble spread for probabilistic forecasting. Another approach is Spherical DYffusion \citep{cachayDYffusionDynamicsinformedDiffusion2023}, a diffusion-based model explicitly designed as a probabilistic atmospheric climate model emulator. cBottle, introduced by \citep{brenowitzClimateBottleGenerative2025}, a generative diffusion model that successfully learns the short-term, instantaneous dynamics of weather and climate when trained on multiple atmospheric datasets, demonstrating the potential for generating realistic atmospheric states. \citep{brenowitzClimateBottleGenerative2025} also trains a version of the model that allows it to capture coherent temporal states. 

A major advance in auto-regressive climate modeling is seen in ACE and its successor, ACE2 (\cite{watt-meyerACEFastSkillful2023,watt-meyerACE2AccuratelyLearning2025}). ACE2 is a deterministic emulator capable of generating stable sequential atmospheric states for up to 1000 years. However, its exclusive focus on the atmosphere omits the essential dynamics of the coupled ocean–atmosphere system. ACE2 is provided CO2, sea ice and SST as boundary conditions, and also lacks a realistic response to different SST trajectories. Additionally, LUCIE3D \citep{guanLUCIE3DThreedimensionalClimate2025} provides an approach aimed at efficiently learning and predicting components of the atmosphere using ERA5 data. LUCIE3D is able to represent both the internal variability and forced response over 30 years of an explicit atmospheric column.

Without an explicit ocean component, atmosphere-only models have a one-way interaction with the ocean, limiting the accuracy and expressivity of major climate drivers (e.g. ENSO). Addressing the need for combining the ocean and atmosphere, several models have emerged. DLESyM \citep{cresswell-clayDeepLearningEarth2025b} is a complex coupled atmosphere-ocean emulator demonstrating the practical feasibility of simultaneously modeling the two major components trained with ERA5 data. The SamudrACE model \citep{duncanSamudrACEFastAccurate2025} complements this by combining the deterministic ocean model Samudra \citep{dheeshjithSamudraAIGlobal2025} with ACE2, resulting in a model that shows effective exchange between the ocean and atmosphere. Similarly, ACE-SOM2 \citep{clarkACE2SOMCouplingML2025} couples ACE2 with a single-layer ocean model (which represents the well-mixed layer) to investigate model response under different forcings, addressing generalization beyond stable climate scenarios. 

In this work, we aim to emulate the IPSL-CM6A-LR climate model at seasonal to decadal timescales with both oceanic and atmospheric state variables. We introduce ArchesClimate, an AI-driven probabilistic emulator specifically designed to overcome the computational bottleneck of exploring the internal variability of the coupled ocean-atmosphere system that aims to learn climatic processes at a monthly temporal resolution. We build on ArchesWeatherGen \citep{couaironArchesWeatherGenSkillfulComputeefficient2026}, a state-of-the-art weather prediction model based on PanguWeather \citep{biPanguWeather3DHighResolution2022}. ArchesWeatherGen provides a computationally efficient solution for ensemble generation. As in ArchesWeatherGen, we use flow matching as our training scheme, a recent generative technique that learns a function to map one distribution to another distribution, usually from a multivariate normal distribution to the target data distribution \citep{lipmanFlowMatchingGenerative2023}. 

We use a dataset from the Decadal Climate Prediction Project (DCPP), a panel of CMIP6 \citep{eyringOverviewCoupledModel2016}. The dataset is made of decadal hindcasts, where the climate model is used to retrospectively forecast the past climate in 10-year chunks, initialized from an observed past state. It comprises ensembles of 10 members with a duration of 10 years, starting every year between 1960 and 2015. This dataset provides us with many samples per year, which allows us to efficiently learn the dynamics of IPSL-CM6A-LR over decadal scales. 

We target a monthly resolution to enhance the computational efficiency of the emulation process. This decision draws upon the theoretical framework that the large-scale ocean and atmosphere coupling operate at a timescale larger than 1 month, while fast weather processes can be treated as a stochastic forcing \citep{hasselmannStochasticClimateModels1976}.

% This decision also draws upon the theoretical framework that long-term system dynamics are driven by large-scale, low-frequency variability \citep{hasselmannStochasticClimateModels1976,heldProbingFastSlow2010}. Therefore, we prioritize the representation of these slower modes by using a monthly timestep.

% We distinguish between an emulator of the input/output functionality of the climate model versus an emulator of how a climate model evolves in climate state. The former describes a machine learning model that takes in boundary conditions or forcings and outputs full climate states, replacing the functionality of a climate model. The latter describes a machine learning model that learns dynamics from the output of a climate model, often augmenting the emulated climate model. In this research, we do the latter, and therefore our setting uses data from the IPSL-CM6A-LR for both initial conditions and training data. If we worked directly from the forcings and did not use an initial state generated from IPSL-6CMA-LR, we could instead attempt to replace the functioning of IPSL-6CMA-LR. 

The manuscript is structured as follows: \cref{sec:datasets} and \cref{sec:arch} present the dataset and architecture used in the research, \cref{sec:training} and \cref{sec:sampling} outline how we train and generate data, and \cref{sec:experiments} explains the experiments and their results. \cref{sec:conclusion} discusses conclusions and next steps.

\section{Materials and Methods}\label{sec:methods}
\subsection{Dataset}\label{sec:datasets}
We use outputs from the IPSL-CM6A-LR coupled climate simulation from the DCPP  \citep{boucherPresentationEvaluationIPSLCM6ALR2020,boerDecadalClimatePrediction2016}. The DCPP experiments aims at exploring decadal climate predictability and variability. We use data from \textit{hindcastA}, an ensemble of 10-year hindcasts starting from 1960 to the present, presented in \cite{swingedouwEvaluationTwoVersions2026}. Ensembles are initialized every year on January 1st from 1960-2015 for an associated coupled run using sea surface temperature and salinity anomaly nudging. For example, the dataset contains a 10-member ensemble from 1960-1970 and another 10-member ensemble for 1961-1971. Despite the use of an almost identical initial state for all runs sharing the same start date, the role of the initial condition in these runs is modest and the runs are uncorrelated after 2 to 5 years \citep{swingedouwEvaluationTwoVersions2026}. It was also found that the hindcasts show no dominant drifts. This allows learning the dynamics of the IPSL-CM6A-LR over a 10-year timeframe with significantly more samples than in other MIP-style experiments. Nevertheless, we speculate that similar results would have been found using other CMIP6 experiments. We use monthly outputs that are averaged values, corresponding to a dataset of approximately 70,000 simulated months.  We will refer to the dataset as IPSL-DCPP hereafter.

We select a subset of the available outputs of the IPSL-CM6A-LR (see \cref{table:vars}). We choose to use the ocean heat content at different depths that represents the vertically integrated heat for these depths. We use 10 surface variables, 7 oceanic variables and 7 atmospheric variables with 4 pressure levels (250, 500, 750, 800 hPa). To limit the use of computing resources, we use this subset of variables as a proof of concept to capture different ocean-atmosphere states at a monthly timescale. 

The surface variables include the heat and water flux between the two domains through $\textit{net\_flux}$ (total heat exchange between atmosphere and ocean, see \ref{appendix:netflux}  for a detailed description) and \textit{evspsbl} (evaporation). The resolution of the atmospheric model is 144x143 (lon x lat, roughly 2.5x1.25 degrees), and the resolution of the ocean model is 362x332 (tripolar, curvilinear orthogonal grid, roughly 1 degree). The oceanic variables are re-gridded onto the regular atmospheric grid using a nearest-neighbour interpolation, with grid points over land set to zero. We also use globally averaged atmospheric concentration of four greenhouse gases (\textit{CO2}, \textit{CH4}, \textit{CFC11eq}, \textit{N2O}) and the incoming solar irradiance used as boundary conditions in the DCPP experiments. These time series are downloaded from \textit{input4mips}, a data repository of standardized boundary conditions used in model-intercomparison projects \citep{meinshausenUoMAIMssp370121GHGConcentrations2018}.

\begin{figure}[H]         
\centering
\includegraphics[width=\linewidth]{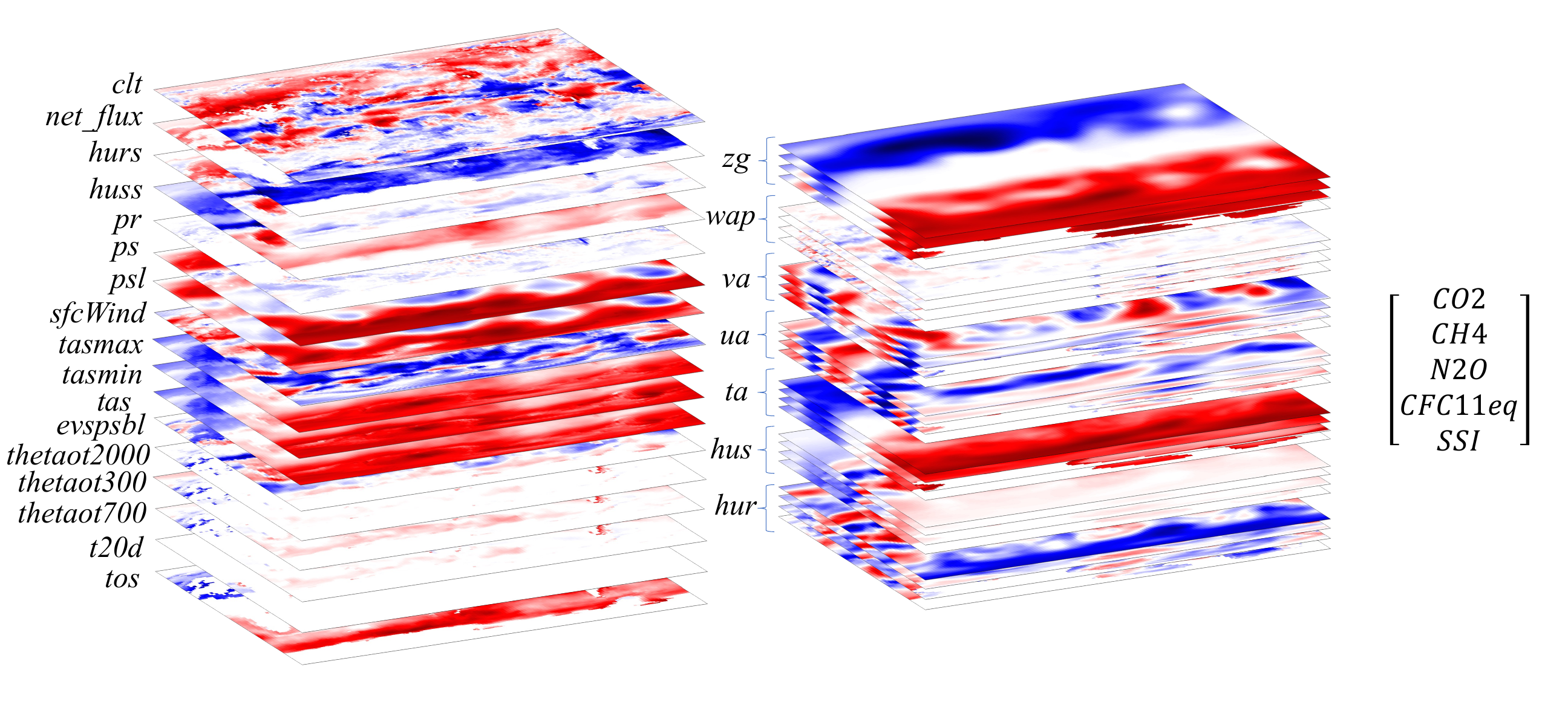}\caption{A visualization of one state ($X_t$) from ISPL-DCPP, with surface (left panel) and atmospheric variables separated from oceanic (right panel) variables. The data used as external forcings in ArchesClimate are shown as a vector to the right. }\label{fig:dcpp_diagram}
\end{figure}

\begin{singlespace}

\begin{table}
% \centering
\caption{Variables in IPSL-DCPP}\label{table:vars}
\begin{tabular}{l l}
\hline
Variable Name & Long Name  \\
\hline
\multicolumn{2}{c}{\textbf{Surface Variables}}\\
\hline\\

\textit{clt} & Total cloud cover percentage  \\
\textit{hurs} & Near-surface relative humidity \\

\textit{huss} & Near-surface specific humidity \\

\textit{pr} & Precipitation  \\

\textit{ps} & Surface air pressure  \\

\textit{tasmax} & Daily maximum near-surface air temperature \\

\textit{tasmin} & Daily minimum near-surface air temperature \\

\textit{tas} & Near-surface air temperature \\

\textit{evspsbl} & Evaporation including sublimation and transpiration\\

\textit{sfcWind} & Near-surface wind speed \\

\textit{net\_flux} & Total positive downward flux between ocean and atmosphere. \\

\textit{psl} & Sea level pressure \\
\\
\multicolumn{2}{c}{\textbf{Ocean Variables}}\\
\hline\\

\textit{thetaot2000} & Vertically-averaged potential temperature at 0-2000m \\

\textit{thetaot700} & Vertically-averaged potential temperature at 0-700m \\

\textit{thetaot300} & Vertically-averaged potential temperature at 0-300m \\

\textit{t20d} & Depth of 20 degree celsius isotherm \\

\textit{tos} & Sea surface temperature \\
\\
\multicolumn{2}{c}{\textbf{Atmospheric Variables at 250, 500, 700, 850 hPa}}\\
\hline\\

\textit{hur} & Relative humidity \\
\textit{hus} & Specific humidity  \\
\textit{ta} & Air temperature \\
\textit{ua} & Eastward wind \\
\textit{va} & Northward wind \\
\textit{wap} & Omega  \\
\textit{zg} & Geopotential height\\
\\
\multicolumn{2}{c}{\textbf{Forcings}}\\
\hline

\textit{CO2} & Atmospheric carbon dioxide (Yearly, Non-spatial) \\
\textit{CH4} & Atmospheric methane (Yearly, Non-spatial) \\
\textit{N2O} & Nitrous oxide (Yearly, Non-spatial) \\
\textit{CFC12eq} & Trichlorofluoromethane (Yearly, Non-spatial) \\
\textit{SSI} & Spectral solar irradiance (Daily, Non-Spatial) \\
\end{tabular}
\end{table}
\end{singlespace}

\subsection{Architecture}\label{sec:arch}
Here we describe the architecture of ArchesClimate. The backbone is similar to ArchesWeatherGen \citep{couaironArchesWeatherGenSkillfulComputeefficient2026}. The architecture consists of a two-part machine learning system for weather forecasting: a deterministic model and a generative model. The deterministic component is built on a 3D Swin U-Net transformer structure based on PanguWeather and employs a novel Cross-Level Attention (CLA) mechanism to efficiently handle interactions within the atmospheric column \citep{biPanguWeather3DHighResolution2022,liuSwinTransformerHierarchical2021,vaswaniAttentionAllYou2017}. The second component, ArchesWeatherGen, is a probabilistic generative model that shares the same architecture and uses a flow matching approach to model the residual weather states, thereby enabling the generation of reliable ensemble forecasts to quantify uncertainty. This combined approach is designed to enhance both the accuracy of the deterministic prediction and the robustness of probabilistic forecasting while being computationally efficient. ArchesWeatherGen operates in a two-stage process to generate its forecasts. First, it uses an ensemble of deterministic models to calculate the most likely future atmospheric conditions, or the ensemble mean, for the next 6-hour timestep. Second, instead of generating a probabilistic forecast of the entire next state, the model focuses its computational power on generating a probabilistic ensemble of residuals. The residual represents the difference between the predicted ensemble mean and the actual state. By focusing on predicting this residual, the model efficiently captures uncertainty and internal variability around the most likely forecast, allowing us to sample different realizations of atmospheric and oceanic dynamics from a fixed base state.

To extend the ArchesWeatherGen approach to the decadal climate prediction domain:  
\begin{itemize}
  \item We implement conditional layer normalization \cite{chenAdaSpeechAdaptiveText2021} for the greenhouse gas and solar irradiance forcings, as given in \cref{table:vars}.
  \item At the cost of increased complexity, we remove axial attention and instead apply self-attention over the fully flattened atmospheric column, allowing the model to learn the relationships between all spatial locations and vertical levels \citep{hoAxialAttentionMultidimensional2019}.
  \item We increase the embedding dimension to 4 times that of ArchesWeatherGen to account for the increased model complexity.
  \item We omit per-variable weighting in the loss function while keeping latitude weighting. Previous studies of weather emulators have included per-variable weighting \citep{biPanguWeather3DHighResolution2022}, but given the different nature of our variable set, which contains ocean and atmospheric variables at a monthly timescale, we cannot reuse those weightings. 
\end{itemize}

Please see \cref{sec:deterministic_ablations} and \cref{sec:generative_rmse_esr} for ablations on these architectural changes. There are also changes to the training of ArchesClimate, which can be found in \cref{sec:training}.

\subsection{Forcings included in ArchesClimate}
The effect of external forcing is as dominant as the effect of internal variability at the decadal timescale \citep{meehlRelativeIncreaseRecord2009}. A representation of the effect of external forcing is needed to emulate the decadal hindcasts. To incorporate forcings in ArchesClimate, we adopt conditional layer normalization following \citep{chenAdaSpeechAdaptiveText2021}. In this approach, each scalar forcing value is first passed through an embedding layer, which produces parameters that modulate the outputs of a standard layer normalization. The forcings are the standardized timeseries of \textit{CO2}, \textit{CH4}, \textit{CFC11eq}, \textit{N2O} and \textit{SSI} as described in \cref{table:vars}.
We apply conditional layer normalization in all transformer blocks of ArchesClimate, enabling the model to integrate conditioning signals at both global and local levels. 

\subsection{Training}\label{sec:training}
\begin{figure}[h!]
\includegraphics[width=\linewidth]{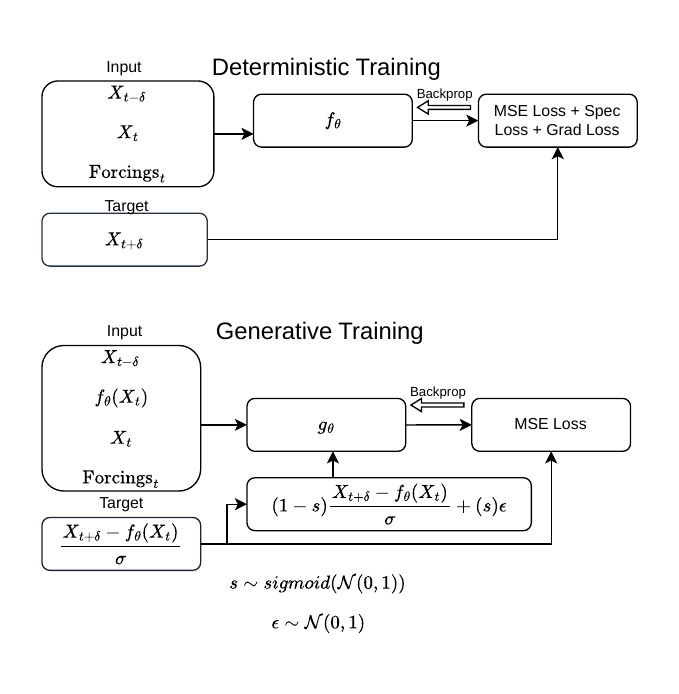}
\caption{Deterministic and Generative training schemes for ArchesClimate. It is necessary to have fully trained $f_{\theta}$ before training $g_{\theta}$. $f_{\theta}$ learns a strong prior of the mean climate, while $g_{\theta}$ learns the residuals of the learned mean climate.}\label{training_scheme}
\end{figure}

Following ArchesWeatherGen, we train both a deterministic and generative model to predict the next state at timestep $t+\delta$, where \textit{t} is a timestep and $\delta$ is one month. A deterministic model, $f_{\theta}$, is trained to predict $X_{t+\delta}$ from $X_{t}$ and $X_{t-\delta}$ where $X$ is a climatic state of IPSL-DCPP (see \ref{fig:dcpp_diagram}). In this climate state, we include the forcings listed in \cref{table:vars} shown as $\text{Forcings}_t$ in \cref{training_scheme} via conditional layer normalization (see previous section). 
We include a spectral and gradient component to the loss of the deterministic model as a way to increase variability. This is a common practice in climate emulation to increase variability \citep{saccardiAssessingGeographicGeneralization2025,lupin-jimenezSimultaneousEmulationDownscaling2025,guanLUCIELightweightUncoupled2025,kochkovNeuralGeneralCirculation2024}. To properly weight the loss functions, we scale the gradient loss and spectral loss by 0.2. Please see \cref{sec:deterministic_ablations} and \cref{sec:generative_rmse_esr} for more details. 
The loss function is as follows:\\
\begin{equation}
\mathcal{L}_{\text{MSE}} = \mathbb{E} \left[ (P_{i,j} - G_{i,j})^2 \right]
\end{equation}
\begin{equation}
\mathcal{L}_{\text{grad}} = \mathbb{E} \left[ \left| (P_{i,j+1} - P_{i,j}) - (G_{i,j+1} - G_{i,j}) \right|^2 \right] + \mathbb{E} \left[ \left| (P_{i+1,j} - P_{i,j}) - (G_{i+1,j} - G_{i,j}) \right|^2 \right]
\end{equation}
\begin{equation}
\mathcal{L}_{\text{PSD}} = \mathbb{E} \left[ \left( \log \left( |\mathcal{F}(P)|^2 + \epsilon \right) - \log \left( |\mathcal{F}(G)|^2 + \epsilon \right) \right)^2 \right]
\end{equation}
\begin{equation}
\mathcal{L}_{\text{total}} = \mathcal{L}_{\text{MSE}} + 0.2*\mathcal{L}_{\text{grad}} + 0.2*\mathcal{L}_{\text{PSD}}
\end{equation}

Where: $P$ is the prediction and subscripts i,k designate the indices for longitude and latitude, $G$ is the ground truth $\mathcal{F}$ denotes the 2D Fourier transform of the predicted image, $|\mathcal{F}|^2$ denotes the Power Spectra Density and $\epsilon$ represents a small regularization factor (1e-10). We will refer to this loss as a composite loss. 

We then train a generative model $g_{\theta}$ to predict the residual of the predicted next state: 
\begin{equation} r_{t+\delta} = \frac{X_{t+\delta} - f_{\theta}(X_{t+\delta}|X_t,X_{t-\delta})}{\sigma}
\end{equation} where $\sigma$ is the standard deviation (averaged spatially and temporally) of the residuals ($X_{t+\delta} - f_{\theta}(X_{t+\delta}|X_t,X_{t-\delta})$) of the training dataset. 

We use flow matching (FM) to train $g_{\theta}$. FM takes known distribution \textit{p} and finds a path of probabilities to an unknown distribution \textit{q} \citep{lipmanFlowMatchingGenerative2023}. This probability path is discretized over 
$S\in\mathbb{N}$ steps, where \textit{S} denotes the total number of discrete time intervals used to approximate the continuous flow from \textit{p} to \textit{q}. In our case, \textit{p} is the Gaussian distribution and \textit{q} is the distribution of the residual of the IPSL-DCPP. Training involves learning $\theta$ such that $g_{\theta}$ predicts $r_{t+\delta}$. The inputs to $g_{\theta}$ are the predicted state of the deterministic model $f_{\theta}(X_t)$, the previous state $X_{t-\delta}$ and a residual noise according to a randomly chosen FM timestep $s \in S$, $(1-s)r_{t+\delta} + s\epsilon$, where $\epsilon$ is noise sampled from a Gaussian distribution. During training, we sample 
\textit{s} from a standard normal distribution $s \sim sigmoid(\mathcal{N}(0,1))$, and use the sigmoid function as done in \cite{esserScalingRectifiedFlow2024a}. See \cref{training_scheme} for a visualization of the process.

At each FM timestep $s$, the probability path is defined by a vector field that assigns a direction and magnitude to each point in our data to move between distributions. $g_{\theta}$ is updated to represent a vector field of the residual $(r_{t+\delta} - \epsilon)$ with the following loss function: 
\begin{equation} \mathcal{L} = \mathbb{E}_{s\in\mathcal{U}(0,1),\epsilon\in\mathcal{N}(0,1)}\|(g_{\theta}(X_t,f_{\theta}(X_{t+\delta}|X_t,X_{t-\delta}),\frac{(1-s)r_{t+\delta} + s\epsilon}{\sigma})- (r_{t+\delta}  - \epsilon)\|_2^2 
\end{equation} We refer to \citep{lipmanFlowMatchingGenerative2023,esserScalingRectifiedFlow2024a} for more details on the flow matching process.

% We do not do out-of-distribution finetuning (training on data outside of the deterministic training dataset) as we have a distribution shift over time that needs to be captured in ArchesClimate. Finetuning could overfit to a particular temporal period. 

The IPSL-DCPP dataset, was carefully split holding out 4 hindcast ensembles (initialized in 1979, 1989, 1999, and 2009) as the validation set and 7 hindcast ensembles (initialized in 1969, and 2010–2015) as the test set, leaving the remaining 44 hindcasts for training. In each case, we use all 10 members. This configuration is justified by the use of distinct initial conditions in each ensemble, which makes the members of each ensemble almost independent after 2 to 5 years. This allows for robust evaluation by comparing a complete 10-member ensemble from a given initialization year against the output of ArchesClimate. In \ref{exp:alternative_train_test} an alternative data split was tested where the simulated year 2010-2020 was held out in the training, and used for evaluation. This method led to overestimated CRPS, presumably due to poor generalization in the years where the external forcing was different. See \cref{sec:training_details} for a list of hyperparameters used. 

\subsection{Inference}\label{sec:sampling}
Once both the deterministic and generative models are trained, ArchesClimate auto-regressively generates sequential states. To generate a state, the model needs to move from Gaussian noise $\epsilon$ to the data distribution of the following step. ArchesClimate takes $M\in \mathbb{N}$ FM steps during inference to go from $r_{t+\delta,0}$ to $r_{t+\delta,S}$ where $M$ is a hyperparameter set at inference time. $M$ is lower than $S$. At each FM inference step $s$, the model outputs $r_{t+\delta,\Delta_{s+1}}$ with ArchesClimate until it reaches $r_{t+\delta,S}$ where $\Delta_s$ consists of evenly spaced values from 0 to S with a step size of $\frac{S}{M}$:
\begin{equation} r_{t+\delta,\Delta_{s+1}} = r_{t_+\delta,\Delta_{s}} + (\Delta_{s+1} - \Delta_{s})g_{\theta}(r_{t_+\delta,\Delta_{s}}|f_{\theta}(X_{t+\delta}|X_t,X_{t-\delta}) \end{equation}
Each step in the sampling process can be interpreted as taking a small step $(\Delta_{s+1} - \Delta_{s})$ in the direction given by the vector field generated with $g_{\theta}$, which is then added to the current state $r_{t_+\delta,\Delta_{s}}$. 
To recreate $X_{t+\delta}$, we combine the output of the deterministic and generative model: 
\begin{equation} \label{eq:next}
X_{t+\delta} = f_{\theta}(X_{t+\delta}|X_t,X_{t-\delta}) + r_{t+\delta,S}
\end{equation}.
 
\begin{figure}[h!]
\includegraphics[width=\linewidth]{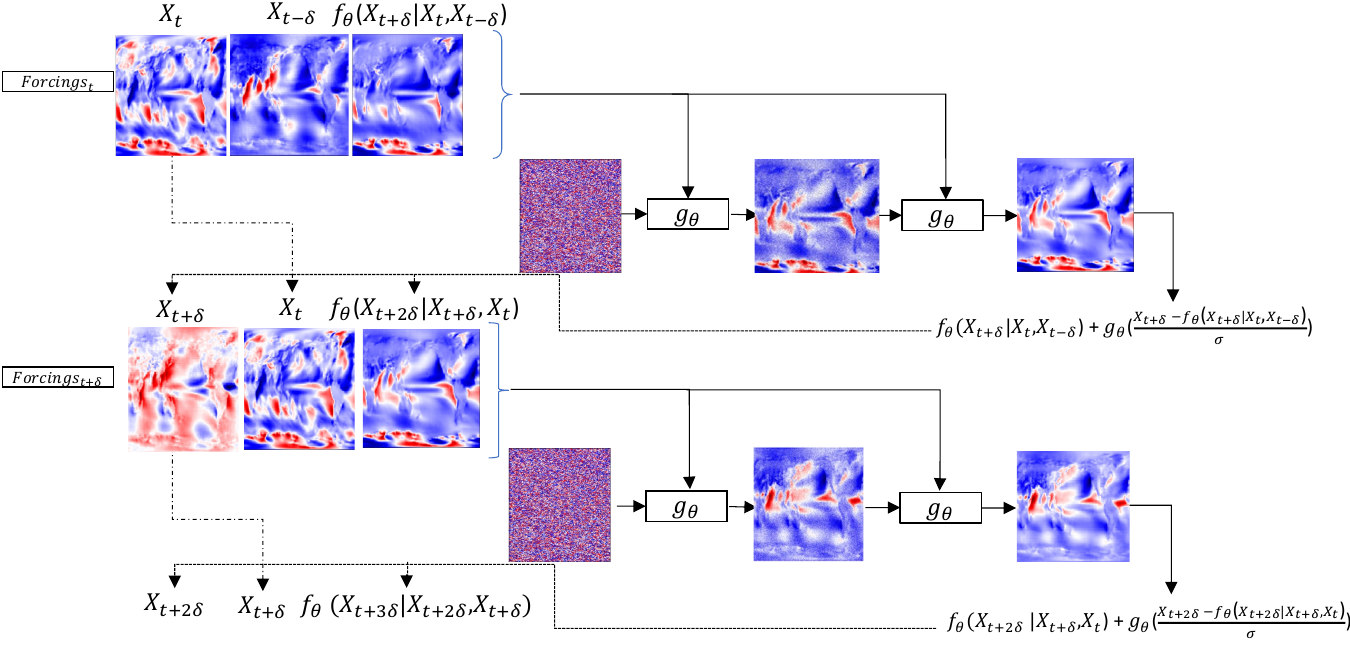}
\caption{Sampling with ArchesClimate. Initial states and noise are given to $g_{\theta}$ and slowly shift from noise to the data distribution over $M$ inference timesteps. The combined result of $f_{\theta}$ and $g_{\theta}$ are then used as input for the following timestep $t$.} \label{fig:sampling_diagram}
\end{figure}

Once ArchesClimate has followed the probability path to the data distribution, it can take $r_{t+\delta,S}$, the output from $g_{\theta}$, and use \cref{eq:next} to build the full next state. This is then passed to $f_{\theta}$ (and subsequently $g_{\theta}$) as input to generate subsequent states. See \cref{fig:sampling_diagram} for a visualization of this process. 
To initialize ArchesClimate for inference, the model begins with the first two monthly states of the IPSL-DCPP hindcasts as an initial state. For instance, January 1969 and February 1969 from an ensemble member of IPSL-DCPP are used to initialized ArchesClimate when emulating the 1969-1979 period.

\subsection{Training Cost Details}\label{sec:training_details}
\begin{table}[h!] \caption{Hyperparameters used in ArchesClimate.}\label{t1} \centering \begin{tabular}{lc} \hline\hline \textbf{Hyperparameter} & \textbf{Value}\\ \hline Learning Rate & $4e^{-4}$ \\ Inference Flow Matching Steps (M) & 12 \\ Training Flow Matching Steps (S) & 1000 \\ Optimizer & AdamW \\ Schedule & Cosine with Warmup \\ Warmup Steps & 5000 \\ Betas ($\beta_1, \beta_2$) & 0.9, 0.98\\ Weight Decay & 0.03 \\ \hline\hline \end{tabular} \end{table}

Quantifying the precise computational speedup of ArchesClimate relative to a traditional Earth System Model (ESM) like IPSL-CM6A-LR remains challenging. However, a rough comparison can be made by estimating the complexity of both approaches and normalizing for hardware differences. 

The IPSL-CM6A-LR model operates on approximately $1.06 \times 10^7$ spatial grid points \citep{acostaComputationalEnergyCost2024}. While the model tracks roughly 750 prognostic variables, the computational load is unevenly distributed. According to \cite{acostaComputationalEnergyCost2024}, the ocean component comprises the majority of the grid ($\approx1 \times 10^7$ points), compared to the atmospheric component ($\approx1 \times 10^6$ points), \citep{boucherPresentationEvaluationIPSLCM6ALR2020}. We therefore estimate the effective parameter space of the IPSL model to be roughly $1.0 \times 10^{10}$ (the number of prognostic variables multiplied by the parameter space). In terms of throughput, IPSL-CM6A-LR completes 12 simulated years per day, consuming approximately 1,900 core-hours per simulated year \citep{acostaComputationalEnergyCost2024}.

In contrast, ArchesClimate uses 45 prognostic variables on a grid of $144 \times 144$ (approx. $10^6$ points). This suggests ArchesClimate is roughly 1/10000th the size of the IPSL model in terms of complexity. Furthermore, we assume here a monthly temporal resolution for ArchesClimate and IPSL, whereas in practice the IPSL model computes prognostic variables at the physical or dynamical model time step (from 5-30 minutes), further widening the computational gap.

To compare costs, we must normalize the hardware usage. Using the approximation that one NVIDIA V100 GPU is roughly equivalent to 200 CPU cores \citep{kimGPUAccelerationMPAS2021}, we can derive a "core-hour" equivalent for the machine learning model. We first need to quantify the core-hours of the training process. A single deterministic model is trained with 6 V100s for 10 hours, which costs 12000 core-hours. The generative model is trained for 20 hours with 6 V100s, which costs 24000 core-hours. During the inference, ArchesClimate emulates 10 years in 12 minutes, or 50 emulated years per wall-clock hour on a single V100 GPU, resulting in a cost of 4 core-hours per simulated year for ArchesClimate. 

Let us consider the case where we want to generate subsequent members for the DCPP project, and want to determine which method is computationally more efficient. This assumes the ensembles generated have a similar size of variable set. The training costs for the generative model are 36,000-core hours (24,000 for the generative model, plus 12,000 for the deterministic model). The IPSL model consumes 19,000 core-hours per 10-year ensemble member. ArchesClimate consumes approximately 40 core-hours per ensemble member. This means that if we want to generate more than one extra ensemble member, it is more computationally efficient to train ArchesClimate and generate each member using this model. This "break-even" point would change depending on the number of variables and temporal resolution. In this comparison that requires more than 10 members, both methods (ArchesClimate and the IPSL) require that the IPSL climate model generate 10 initial ensemble members.

While ArchesClimate is drastically cheaper in absolute terms (4 vs. 1,900 core-hours per simulated year), this efficiency is largely a function of its reduced resolution. If we were to scale ArchesClimate linearly to match the IPSL's parameter space (multiplying the cost by the size factor of ~10000), the theoretical cost would sharply rise. This implies that at equivalent complexity and resolution, the ML approach might be only marginally less expensive than the physics-based GCM. Moreover, if higher temporal resolutions were required, the ML cost could exceed that of the GCM. However, linear scaling may be an overly pessimistic assumption, as ML techniques often leverage latent spaces to improve efficiency beyond brute-force scaling. The primary value of ArchesClimate is its ability to approximate IPSL dynamics, delivering the most useful variables without running the full simulation. Both physics-based and ML models require extensive tuning, adding further complexity to a direct "price tag" comparison. While ML models utilize gradient descent for internal parameter optimization, both GCMs and ML models require an external search phase (tuning and hyperparameter optimization, respectively) to maximize performance. Finally, there is an advantage regarding engineering accessibility: generating subsequent ensemble members using ArchesClimate requires only a single, widely available commodity GPU rather than a HPC machine needed to run the IPSL model. 

\section{Experiments}\label{sec:experiments}
In this section, we describe the setup, evaluation methods and diagnostic tools used in experiments conducted with ArchesClimate. We then describe each experiment and its results.

\subsection{Calculation of Climatology} \label{sec:climatology_calc}
For the experiments that use a deseasonalized state, we remove the seasonal cycle from each dataset. We compute the seasonal average across two decades in the test period (the ensemble starting at 1969 and the ensemble starting at 2010), giving 20 years (240 months) to calculate the climatology. This climatology is removed spatially from the data. We compare the climatology of the generative models in \cref{sec:climatology_analysis}.

\subsection{Baselines} \label{sec:baseline}
To measure how well ArchesClimate captures the dynamics of IPSL-DCPP, we consider the following baselines.

\textbf{Held-out IPSL-DCPP Ensemble} For each target ensemble, we select 5 members of the 10 members to serve as a baseline. We then compare both the remaining 5 ensemble members of IPSL-DCPP and a 5-member ArchesClimate ensemble to the baseline.

\textbf{MESMER-M}
As one of our baselines, we utilize MESMER-M \citep{nathMESMERMEarthSystem2022,beuschEmulatingEarthSystem2020}, a well-established statistical climate emulator that generates spatially explicit monthly temperature fields by decomposing the climate signal into mean and residual components. It is based on pattern scaling for the mean component, and a first order autoregressive process for the residual part. While the foundational MESMER framework operates at an annual scale \citep{beuschEmissionScenariosSpatially2022}, MESMER-M extends this to monthly resolution \citep{nathMESMERMEarthSystem2022}. 

For our experimental setup, we calibrated MESMER-M using the same data for the training period as ArchesClimate. Anomalies were calculated relative to the 1960-1970 reference period. During the evaluation phase, we provided MESMER-M with the spatially-averaged annual mean surface air temperatures from IPSL-DCPP to predict the monthly temperature of the held-out periods. Further technical details regarding the MESMER-M architecture can be found in \citep{nathMESMERMEarthSystem2022}.

\textbf{DiffESM}
As another baseline, we utilize DiffESM \citep{bassettiDiffESMConditionalEmulation2024}, a deep learning climate emulator designed to conditionally generate spatio-temporally consistent monthly temperature fields from coarser decadal data. Unlike traditional statistical methods, DiffESM is built on a Denoising Diffusion Probabilistic Model (DDPM) that maps Gaussian noise to physically realistic weather sequences. Its core architecture relies on a 3D U-Net that splits downsampling and upsampling tasks into distinct spatial and temporal convolutions. The original model receives monthly averages to predict daily data. To train the DiffESM model, we use the same training dataset split as provided to ArchesClimate. We provide DiffESM with the decadal average of air surface temperature, and the model targets the entire monthly 10-year sequence of an ensemble member from IPSL-DCPP. By doing so, we align the goal of DiffESM with ArchesClimate and change the temporal scale of the original research. The rest of the training parameters are the same as \cite{bassettiDiffESMConditionalEmulation2024}. By conditioning the generative process on decadal averages and monthly labels, the model captures non-stationary frequencies while matching true regional mean constraints.

\subsection{Evaluation Metrics and Diagnostic Tools}\label{sec:eval_metrics}

\textbf{Root Mean Squared Error} (RMSE) is used to evaluate the deterministic accuracy of the emulated ensemble mean against the target ensemble mean. Following the spatial weighting protocol in \cite{couaironArchesWeatherGenSkillfulComputeefficient2026}, the RMSE is calculated by first taking the mean across all ensemble members for both the prediction and the target at each grid point and timestep. The squared error between these ensemble means is then latitude-weighted to account for grid-cell area convergence at the poles, averaged spatially, and subsequently averaged over all temporal samples:

% \begin{equation}\text{RMSE} = \frac{1}{N} \sum_{n=1}^{N} \sqrt{\sum_{i,j} w_{ij} \left( \overline{x}^{(n)}{ij} - \overline{y}^{(n)}{ij} \right)^2}\end{equation}

% where $w_{ij}$ represents the normalized latitude weights ($\sum_{i,j} w_{ij} = 1$), and $\overline{x}^{(n)}_{ij} = \frac{1}{M}\sum_{m=1}^{M} x^{(n,m)}_{ij}$ and $\overline{y}^{(n)}_{ij} = \frac{1}{M}\sum_{m=1}^{M} y^{(n,m)}_{ij}$ denote the emulated and target ensemble means at sample $n$ and grid point $(i,j)$, respectively.

\textbf{Ensemble Spread Ratio} (ESR) is defined as the ratio of the predicted ensemble standard deviation to the target ensemble standard deviation, both computed across members at each grid point and then averaged spatially and over time:

\begin{equation}
\text{Spread Ratio} = \frac{\sqrt{\frac{1}{N}\sum_{n=1}^{N}\sum_{i,j} w_{ij}\text{Var}_{m}\left(x^{(n,m)}_{ij}\right)}}{\sqrt{\frac{1}{N}\sum_{n=1}^{N}\sum_{i,j} w_{ij}\text{Var}_{m}\left(y^{(n,m)}_{ij}\right)}}
\end{equation}

where $\text{Var}_m$ denotes the variance across the $M$ ensemble members, $N$ is the total number of samples, and $w_{ij}$ represents the normalized latitude weights that sum to 1 across all grid points ($\sum_{i,j} w_{ij} = 1$). The numerator uses the predicted ensemble members and the denominator uses the held-out IPSL target ensemble members. A value of 1 indicates that the predicted ensemble has the same spread as the IPSL internal variability; values below 1 indicate the model is underdispersed relative to the target ensemble.
% \textbf{Continuous Ranked Probability Score} (CRPS) is a metric used to evaluate the accuracy of probabilistic forecasts by measuring the difference between the predicted cumulative distribution function and the ensemble mean of the target 5-member ensemble. We use the same implementation of CRPS as in \cite{raspWeatherBench2Benchmark2024} and is latitude-weighted. 

% \textbf{Spread-Skill Ratio} (SSSR) is a metric comparing the internal variance and the mean error of an ensemble forecast. It is defined as the ratio of the ensemble spread (standard deviation among members) to the Root Mean Square Error (RMSE) of the ensemble mean against a target verification. A perfectly calibrated, reliable ensemble yields an SSSR of $1$. Values below unity signify an under-dispersed (overconfident) forecast, while values above unity indicate an over-dispersed (underconfident) forecast. 

% However, evaluating finite sub-ensembles introduces a structural sampling bias; when verifying an $M$-member forecast against an $N$-member target, the expected baseline ratio mathematically shifts above unity to $\sqrt{(M+N)/N}$. In our framework ($M=5, N=5$), the baseline expectation becomes $\sqrt{2} \approx 1.41$. For highly chaotic climate variables with strong internal variability, this baseline ratio can naturally approach or exceed $2.0$ (see \cref{sec:generative_ablation}). Consequently, ArchesClimate is evaluated not against an idealized unit score, but against the true elevated SSSR baseline computed from the original IPSL-DCPP data.

\textbf{Rank Histograms} are a diagnostic tool used to evaluate the consistency or calibration of an ensemble forecast by comparing it to another ensemble \citep{hamillInterpretationRankHistograms2001}. Using rank histograms, we evaluate whether an ensemble member generated with ArchesClimate can be interchanged with any member of the IPSL-DCPP ensemble. To calculate the rank of an ensemble member, we take the pixel-by-pixel values of the ensemble member and compare them to a target ensemble. We calculate the rank of the pixel value of the ensemble member compared to the rest of the target ensemble members at that pixel. We then take an average of the rank of each pixel across space and time. The histograms show the normalized frequency of the emulated ensemble member at that rank. We normalize the frequency by taking the mean and standard deviation of all frequencies across space and time. If the generated ensemble member can be interchanged with any member of an IPSL-DCPP ensemble, the histogram will be flat, showing an even distribution across all ranks.

\textbf{Temporal Power Spectra} (TPS) describes how the variance (or power) of a signal that changes with time is distributed among components that oscillate at specific rates (temporal frequencies, measured in cycles per unit time). This highlights the strength of physical signals at varying timescales. 

We compute the TPS by first calculating the temporal Fourier transform of the time series at each spatial location for an individual ensemble member. For each member, the magnitude of the transform is squared to get the power, latitude-weighted to account for the convergence of grid cells toward the poles, and averaged spatially. Finally, this spatially averaged power spectrum is averaged across all ensemble members.

We formalize this definition below: 

Given a time series for an ensemble member $m$ at a specific grid point $x^{(m)}(n, \mathrm{lat}, \mathrm{lon})$, where $T := 120$ months (10 years), $n = 0, \dots, 119$, and the spatial bounds are defined by:
\begin{equation*}
\mathrm{lat} = 1, \dots, N_y,\quad
\mathrm{lon} = 1, \dots, N_x,\quad
\mathcal{K} = \left\{\, k \;\middle|\; 1 \le k \le \left\lfloor \frac{T}{2} \right\rfloor = 60 \,\right\}
\end{equation*}

The discrete Fourier transform for member $m$ is given by:
\begin{equation}
X^{(m)}(k,\mathrm{lat},\mathrm{lon}) = \sum_{n=0}^{119} x^{(m)}(n,\mathrm{lat},\mathrm{lon}) \, e^{- i 2 \pi k n / 120}
\end{equation}

Let $w_{\mathrm{lat}} = \cos(\phi_{\mathrm{lat}}) / \sum_{\mathrm{lat}'}\cos(\phi_{\mathrm{lat}'})$ represent the normalized latitude weights. The ensemble-averaged, spatially-weighted Power Spectral Density ($\overline{\mathrm{PSD}}$) for frequency $k \in \mathcal{K}$ across $M$ ensemble members is:
\begin{equation}
\overline{\mathrm{PSD}}(k) = \frac{1}{M \cdot N_x} \sum_{m=1}^{M} \sum_{\mathrm{lat}=1}^{N_y} \sum_{\mathrm{lon}=1}^{N_x} w_{\mathrm{lat}} \left|X^{(m)}(k, \mathrm{lat}, \mathrm{lon})\right|^2
\end{equation}

\subsection{Deterministic Model Ablations}\label{sec:deterministic_ablations}
We evaluate the architectural choices detailed in \cref{sec:arch}, assessing the performance of the deterministic component. \cref{table:deterministic_ablations} examines the deterministic model's sensitivity to embedding size, the inclusion of a composite loss term (spectral and grad), the use of axial attention, and the use of conditional layer normalization. All model predictions are derived from a 5-member ensemble. A full grid search of the ablated models is not possible due to computational restraints, but we provide here a subset that permits assessment of the architectural choices. Notably, the 3D configuration with an embedding size of 768 was excluded from our suite due to training instabilities. The metrics are calculated on the decades 1969-1979 and 2010-2020. 

Several key insights emerge from the deterministic ablation results presented in \cref{table:deterministic_ablations}. First, omitting the influence of external greenhouse gas (GHG) and solar forcings through the normalization layer markedly degrades model performance as shown by decades of research on decadal hindcast \citep{meehlRelativeIncreaseRecord2009,vanoldenborghDecadalPredictionSkill2012}. Although the forced climate response remains small over a ten-year horizon, these external forcings provide crucial physical constraints and on a 1 to 10 year time horizon guide the model's trajectory. Second, axial attention does not outperform 2D attention baseline at an equivalent embedding size, performing only better at \textit{pr}, \textit{hur} and \textit{wap}. Presumably, the vertical relationships between levels is more readily learnable when the layers are flattened along the channel dimension with a high-dimensional embedding space. The two tests with an embedding size of 768 result in a RMSE lower than their respective models at an embedding size of 192. The models with an composite loss component (S/NS) have a lower RMSE than their counterparts without a composite loss component. We hypothesize that the inclusion of the composite loss component in the deterministic model will improve the variance of the generative model, as a more accurate deterministic model will allow the the generative model to focus its capacity on modeling the internal variability rather than fixing structural deficits. To address this, both the 2D-768-NCL-C and 2D-768-CL-C architectures are retained as downstream baselines for the generative models explored in the following section.

\begin{table}[h!]
\centering
\resizebox{\textwidth}{!}{%
\begin{tabular}{l|ccccccc}
\toprule
\textbf{Variable} & \textbf{2D-768-CL-NC} & \textbf{3D-192-CL-C} & \textbf{2D-192-CL-C} & \textbf{2D-768-NCL-C} & \textbf{2D-192-NCL-C} & \textbf{2D-768-CL-C} & \textbf{IPSL} \\
\midrule
\textit{tas} & 0.7893 & 0.7637 & 0.5971 & 0.7154 & 0.8345 & \textbf{0.5444} & 1.097 \\
\textit{tas} (land) & 0.9654 & 0.9204 & 0.6982 & 0.7884 & 0.904 & \textbf{0.6409} & 1.337 \\
\textit{tas} (yearly) & 0.8078 & 0.6861 & 0.4164 & 0.4526 & 0.7675 & \textbf{0.2907} & 0.9413 \\
\textit{tas} (land, yearly) & 0.9966 & 0.6182 & 0.3815 & 0.427 & 0.6818 & \textbf{0.2866} & 1.087 \\
\textit{thetaot2000} & 0.1901 & 0.3826 & 0.2077 & 0.2216 & 0.4965 & \textbf{0.1569} & 0.2201 \\
\textit{tos} & 0.5401 & 0.608 & 0.4826 & 0.4949 & 0.6942 & \textbf{0.3507} & 0.6628 \\
\textit{psl} & 108.92 & 113.76 & 106.32 & 144.75 & 156.31 & \textbf{99.99} & 179.17 \\
\textit{net\_flux} & 8.973 & 11.65 & 10.19 & 10.29 & 12.27 & \textbf{8.25} & 12.62 \\
\textit{pr} & 6.67e-06 & 5.85e-06 & 6.20e-06 & 6.56e-06 & 6.60e-06 & \textbf{5.82e-06} & 9.49e-06 \\
\textit{hur} 700hPa & 2.908 & \textbf{2.603} & 2.742 & 2.911 & 2.868 & 2.64 & 3.999 \\
\textit{ta} 700hPa & 0.7206 & 0.6605 & 0.5454 & 0.5939 & 0.6965 & \textbf{0.4256} & 0.9938 \\
\textit{wap} 700hPa & 0.01049 & 0.01043 & 0.01047 & 0.0107 & 0.01119 & \textbf{9.70e-03} & 0.01641 \\
\textit{zg} 700hPa & 11.36 & 10.86 & 10.09 & 13.85 & 14.92 & \textbf{9.245} & 17.87 \\
\bottomrule
\end{tabular}%
}

\caption{Results from ablations applied to the deterministic model indicated by the CRPS. The first line provides a name of the test. In the name, 2D/3D indicates the use of axial attention. The following number is embedding size. CL/NCL indicates the use of an composite loss component. Lastly, C/NC shows the use of conditional layer normalization. The column IPSL shows the RMSE for the held-out 5-member ensemble }
\label{table:deterministic_ablations}
\end{table}

\subsection{RMSE and ESR of Generative Model Ablations}\label{sec:generative_rmse_esr}

In this section, we expand our exploration of model selections by testing the ensemble size. To complement the  comparison with MESMER-M and DiffESM, the metrics are also computed on annual mean data. \cref{table:generative_rmse} and \cref{table:generative_esr} shows the RMSE and Ensemble Spread Ratio (ESR) for each test. We adopt the naming convention established in \cref{sec:deterministic_ablations}, appending "E" and "S" to denote ensemble ($n=4$) and single-member ($n=1$) configurations, respectively. Here, "NCL" (No Composite Loss) and "CL" (Composite Loss) specify the configuration of the underlying deterministic model, as the composite loss is applied exclusively to the deterministic component. As with the deterministic ablations, we show a subset of the possible ablations to highlight the significant differences in performance. The metrics are calculated on the decades 1969-1979 and 2010-2020.

First, to assess whether integrating a composite loss into the deterministic model improves downstream performance, we compare 2D-S768-CL against 2D-S768-NCL. We focus this comparison on variants with a single deterministic model, as using an ensemble averages the deterministic state and would counteract the effects of the composite loss. We can see in \cref{table:generative_rmse} and \cref{table:generative_esr} including the composite loss improves ESR but also increases RMSE. This behavior stems from the deterministic base model's objective function. By utilizing a composite loss rather than MSE, the base model is forced to preserve high-frequency, energetic features. When this state is passed as conditioning to the generative model, these energetic features are retained. This increases the ensemble spread and subsequently raises the RMSE.

Second, regarding ensemble size, configurations utilizing a deterministic model ensemble exhibit a lower ESR but a higher RMSE. This behavior also suggests that a more averaged (or one with less high frequencies) deterministic state restricts the generative model from learning the proper spread of the target ensemble. Consequently, we observe two methods that improve variability: incorporating composite loss terms into the deterministic model, and using a single deterministic model.

Finally, examining model capacity, the 2D-E192-S configuration exhibited the lowest RMSE for surface variables. It is likely that this reduced capacity limited the model's ability to learn the relationships between atmospheric variables, thereby degrading its performance in those areas. When comparing small and large embedding sizes, the smaller model successfully captures surface dynamics but struggles with atmospheric variables and precipitation. In contrast, the larger embedding size yields highly accurate results across atmospheric, surface, and ocean variables. Lastly, the 3D model performs similarly to the other large embedding size models in terms of both RMSE and ESR.

We also contextualize these results against MESMER-M and DiffESM. We can see that DiffESM struggles to accurately replicate the monthly seasonal cycle as it has an extremely large RMSE for \textit{tas}. This is not the same for the yearly values, and DiffESM performs well at the yearly timestep. Both DiffESM and MESMER-M have higher ESR  than the ArchesClimate models at the yearly timestep, and MESMER-M has a higher ESR at the monthly timestep as well. As a tool that generates the entire temporal sequence simultaneously, DiffESM inherently compromises on fine-grained temporal consistency but excels at rapidly generating annual-scale variability. MESMER-M similarly demonstrates strong performance on monthly metrics, which is expected given that it is explicitly constrained by the global mean temperature. While these models are not auto-regressive, they provide a rigourous baseline for internal variability.

We also point out that the RMSE is almost always lower than the held-out 5-member IPSL ensemble. While accuracy of the model is an important aspect, ensemble spread, interchangeability, and a comparable power spectra all can indicate a good emulation. We will investigate the latter two properties in the following sections.

% we determine if the emb_size has an impact on training, using emb_dim of 192 (as in archesweather) or 768 (4x greater).

% does emb_size matter? 
% emb_size=768,ens_size=4,spectral_det=True,2d_attention flow_ensemble_large_0
% emb_size=192,ens_size=4,spectral_det=True,2d_attention flow_ensemble_small_0

% does spectral in det matter (ens_size = 1)? 
% emb_size=768,ens_size=1,spectral_det=True,2d_attention flow_ensemble_single_small_0
% emb_size=768,ens_size=1,spectral_det=False,2d_attention flow_ensemble_single_not_spectral_1

% does spectral in det matter (ens_size = 4)? 
% emb_size=768,ens_size=4,spectral_det=True,2d_attention flow_ensemble_large_0
% emb_size=768,ens_size=4,spectral_det=False,2d_attention flow_ensemble_2

% does ensemble size matter? 
% emb_size=768,ens_size=4,spectral_det=True,2d_attention flow_ensemble_large_0
% emb_size=768,ens_size=1,spectral_det=True,2d_attention flow_ensemble_single_small_0

% does axial attention matter (big_emb, small_emb)
% emb_size=192,ens_size=1,spectral_det=True,3d_attention flow_3d_prior_small_1
% emb_size=768,ens_size=1,spectral_det=True,3d_attention flow_3d_prior_1
% emb_size=192,ens_size=4,spectral_det=True,3d_attention flow_3d_prior_small_ensemble_1

\begin{table}[h!]
\centering
\resizebox{\textwidth}{!}{%
\begin{tabular}{l|ccccccccc}
\toprule
\textbf{Variable} & \textbf{2D-E192-CL} & \textbf{2D-E768-CL} & \textbf{2D-E768-NCL} & \textbf{2D-S768-CL} & \textbf{2D-S768-NCL} & \textbf{3D-S768-CL} & \textbf{diffesm} & \textbf{MESMER-M} & \textbf{IPSL} \\
\midrule
\textit{tas} & \textbf{0.6525} & 0.6975 & 0.7366 & 0.9734 & 0.7977 & 0.8121 & 6.701 & -- & 1.097 \\
\textit{tas} (land) & \textbf{0.8115} & 0.8724 & 0.9073 & 1.197 & 1.001 & 0.9971 & 9.805 & 1.426 & 1.337 \\
\textit{tas} (yearly) & \textbf{0.3368} & 0.4253 & 0.4481 & 0.5256 & 0.4491 & 0.557 & 0.5457 & -- & 0.9413 \\
\textit{tas} (land, yearly) & \textbf{0.3512} & 0.4205 & 0.4734 & 0.5203 & 0.4287 & 0.5118 & 0.7478 & 0.5164 & 1.087 \\
\textit{thetaot2000} & \textbf{0.1395} & 0.1518 & 0.1444 & 0.1878 & 0.1684 & 0.1619 & -- & -- & 0.2201 \\
\textit{tos} & \textbf{0.4085} & 0.4364 & 0.4161 & 0.5909 & 0.4818 & 0.5255 & -- & -- & 0.6628 \\
\textit{psl} & \textbf{143.59} & 157.83 & 143.95 & 175.46 & 163.85 & 177.57 & -- & -- & 179.17 \\
\textit{net\_flux} & \textbf{11.30} & 11.39 & 11.61 & 16.26 & 13.10 & 13.35 & -- & -- & 12.62 \\
\textit{pr} & 9.10e-06 & \textbf{8.42e-06} & 9.98e-06 & 1.56e-05 & 1.11e-05 & 1.21e-05 & -- & -- & 9.49e-06 \\
\textit{hur} 700hPa & 4.717 & \textbf{3.618} & 3.736 & 5.617 & 4.248 & 4.524 & -- & -- & 3.999 \\
\textit{ta} 700hPa & 0.7333 & \textbf{0.6197} & 0.6267 & 0.8606 & 0.7044 & 0.7176 & -- & -- & 0.9938 \\
\textit{wap} 700hPa & 0.01936 & \textbf{0.01503} & 0.01583 & 0.02243 & 0.01753 & 0.01819 & -- & -- & 0.01641 \\
\textit{zg} 700hPa & 15.45 & 14.54 & \textbf{13.64} & 16.75 & 15.20 & 16.07 & -- & -- & 17.87 \\
\bottomrule
\end{tabular}%
}
\caption{Results from ablations applied to the deterministic model indicated by the RMSE. The first line provides a name of the test. In the name, 2D/3D indicates the use of axial attention. The following number is embedding size. CL/NCL indicates the use of a composite loss component. The column IPSL shows the RMSE for the held-out 5-member ensemble. }
\label{table:generative_rmse}
\end{table}

\begin{table}[h!]
\centering
\resizebox{\textwidth}{!}{%

\begin{tabular}{l|cccccccc}
\toprule
\textbf{Variable} & \textbf{2D-E192-CL} & \textbf{2D-E768-CL} & \textbf{2D-E768-NCL} & \textbf{2D-S768-CL} & \textbf{2D-S768-NCL} & \textbf{3D-S768-CL} & \textbf{diffesm} & \textbf{MESMER-M} \\
\midrule
\textit{tas} & 0.4817 & 0.5088 & 0.5963 & \textbf{0.8661} & 0.8148 & 0.7745 & 3.931 & -- \\
\textit{tas} (land) & 0.4749 & 0.5017 & 0.5904 & 0.8638 & 0.8156 & 0.7763 & 4.385 & \textbf{0.9442} \\
\textit{tas} (yearly) & 0.4457 & 0.4957 & 0.5511 & 0.7168 & \textbf{0.764} & 0.763 & 0.7356 & -- \\
\textit{tas} (land, yearly) & 0.4426 & 0.491 & 0.5672 & 0.7544 & 0.7867 & 0.7903 & \textbf{0.9267} & 0.8073 \\
\textit{thetaot2000} & 0.542 & 0.5845 & 0.693 & \textbf{0.9598} & 0.8348 & 0.8419 & -- & -- \\
\textit{tos} & 0.564 & 0.6161 & 0.6709 & \textbf{0.9398} & 0.8546 & 0.8494 & -- & -- \\
\textit{psl} & 0.4142 & 0.4679 & 0.4811 & 0.6713 & \textbf{0.6991} & 0.6906 & -- & -- \\
\textit{net\_flux} & 0.5703 & 0.5883 & 0.6474 & \textbf{1.08} & 0.8566 & 0.8876 & -- & -- \\
\textit{pr} & 0.71 & 0.5982 & 0.8641 & 1.604 & \textbf{1.04} & 1.153 & -- & -- \\
\textit{hur} 700hPa & 1.043 & 0.6283 & 0.6982 & 1.252 & 0.8998 & \textbf{0.9593} & -- & -- \\
\textit{ta} 700hPa & 0.8738 & 0.5157 & 0.5925 & \textbf{0.8905} & 0.8426 & 0.8158 & -- & -- \\
\textit{wap} 700hPa & 1.066 & 0.6636 & 0.7529 & 1.264 & 0.9137 & \textbf{0.9534} & -- & -- \\
\textit{zg} 700hPa & 0.6794 & 0.4532 & 0.4602 & 0.6145 & \textbf{0.6855} & 0.6577 & -- & -- \\
\bottomrule
\end{tabular}%
}
\caption{Results from ablations applied to the deterministic model indicated by the Ensemble Spread Ratio (ESR). The first line provides a name of the test. In the name, 2D/3D indicates the use of axial attention. The following number is embedding size. CL/NCL indicates the use of a composite loss component. }
\label{table:generative_esr}
\end{table}

\subsection{Climatological Biases in ArchesClimate}\label{sec:climatology_analysis}
We examine the climatology of the generative models to understand their respective biases (see \cref{sec:climatology_calc} for calculation details). By calculating the monthly climatology of the test period, an average of the periods 1969–1979 and 2010–2020, we can identify each model's seasonal bias.

\begin{figure}[H]
\includegraphics[width=1\textwidth]{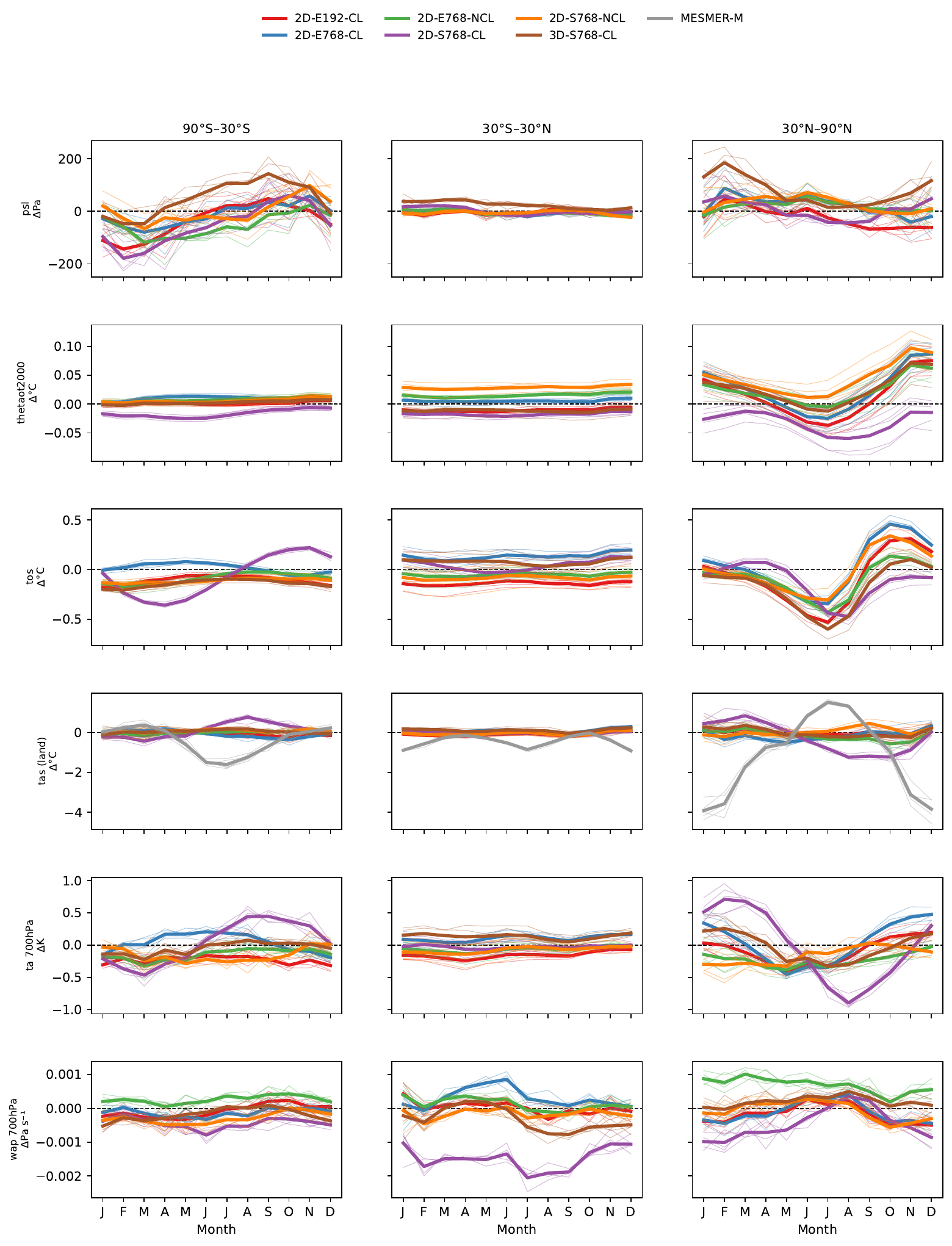}
{\caption{Comparison of climatologies computed over the test period 1969-1979 and 2010-2020. The dotted line represents the ensemble mean, with the thin lines representing the difference from the IPSL ensemble mean to each ensemble member. The different colors correspond to different generative models, as defined in \cref{sec:generative_rmse_esr}. } 
\label{fig:climatology}}
\end{figure}

As shown in \cref{fig:climatology}, there is almost no seasonal bias in the Tropics (30°S–30°N). 2D-S768-CL is the most biased model, exhibiting significant bias in \textit{thetao2000}, \textit{wap}, \textit{tos}, and \textit{ta}, with the latter two displaying inconsistent bias across the year. This indicates that this model struggled to capture the seasonal cycle. In contrast, the equivalent model without the composite loss shows no seasonal offset in its bias. This model also exhibits significant biases in the northern region for \textit{tos}, an area known for high variability over decadal timescales \citep{mignotDecadalPredictionSkill2016}. Additionally, DiffESM clearly misses the seasonal cycle and is therefore excluded from the analysis to improve readability. We can see several degrees of bias in MESMER-M in the Northern region, possibly due to an artifact of linearly predicting monthly temperature values from annual values. 

There is no agreement in bias across the models; those with deterministic ensembles demonstrate lower bias during the SON/DJF months, but greater bias during the MAM/JJA months. This bias could potentially be corrected through fine-tuning with spatial weighting based on existing errors. We analyze the climatology and bias in this section, as the remainder of the analysis uses data with the climatology, and therefore the mean bias, removed.

\subsection{Interchangeability using Rank Histograms}\label{exp:interchangeability}

To evaluate whether ensemble members generated by ArchesClimate are exchangeable with those of IPSL-DCPP, we conducted a rank histogram analysis. We compared a 5-member ArchesClimate ensemble against a 5-member IPSL-DCPP ensemble for two decades in the test period, one initialized at 1969 and one initialized at 2010. Each rank is calculated seperately for each decade before averaging the results. We remove the climatology from each of the ensembles before calculating the rank histograms. This gives us many samples to have a clear picture on the interchangeability of the ensemble members generated by ArchesClimate. See \cref{sec:baseline} for a detailed methodology on rank histograms. We compare the results to the 5-member IPSL held-out ensemble, and the DiffESM and MESMER-M baselines where possible. 

First, we note that the 5-member ensemble of the IPSL-DCPP held-out data produces a flat rank histogram for all variables except \textit{thetaot2000}, which has a slight cold bias. This confirms its exchangeability and validates it as a rigorous baseline for our model comparison. Because our evaluation framework ranks individual predicted members against the distribution of the IPSL-DCPP truth ensemble, the interpretation of the rank histograms is inverted relative to standard evaluations. For instance, in \textit{tas (land)}, the diffESM baseline exhibits distinct U-shaped histogram. This indicates that its generated ensembles are highly over-dispersive, confirming it's inability to properly represent monthly dynamics. This over-dispersion means that it frequently produce extreme local temperature values that fall completely outside the upper and lower bounds of the natural IPSL-DCPP spread.

ArchesClimate exhibits varying degrees of dispersion depending on the model. For \textit{psl}, driven by atmospheric dynamics, ArchesClimate exhibits an n-shaped (convex) histogram, indicating its predictions are underdispersive. This underdispersion aligns with the findings in the preceding section, which showed that across all models, the ESR of \textit{psl} is low. We can also see that for the models with a single deterministic model, they generally have improved interchangeability compared to the other models due to an increased dispersion. They are also the only models that are over dispersive for \textit{tos}. For the atmospheric variables \textit{ta} and \textit{ta}, the model without the composite loss component has almost no bias, exhibiting good exchangeability. This provides a clearer picture for which method improves variability - while both the generative model with the composite loss component and the generative model with the single determinstic model both exhibited improved ESR, here we see that 2D-S768-NCL is significantly more exchangaeble for \textit{wap}, and comparable to 2D-S768-CL for the rest of the variables. It is likely therefore that the composite loss maybe introduce too much unstructured noise into the model. It is interesting however to note that 3D-S768-CL performs very similarly to 2D-S768-NCL. See the following section for further comparison of these two models.

\begin{figure}[H]
    \centering
    \includegraphics[width=\textwidth]{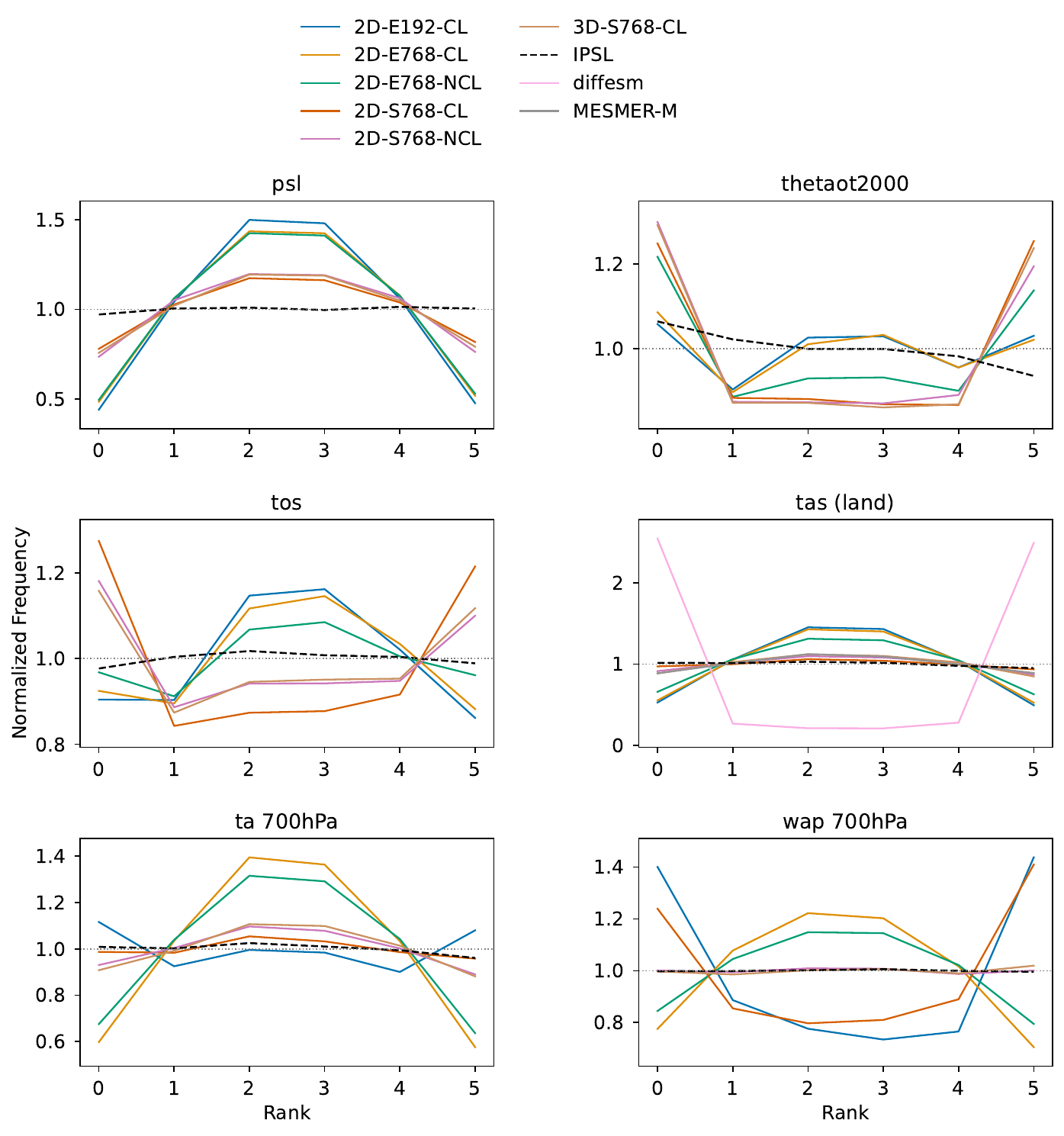}
    \caption{Rank histograms for ArchesClimate for the test period 1969-1979 and 2010-2020, showing the normalized frequency of the rank of each ArchesClimate member within a 5-member ensemble of IPSL-DCPP. Values closer to the black dotted line indicate better interchangeability with IPSL-DCPP.}
    \label{fig:headline_vars_rank_hist}
\end{figure}

\subsection{Comparison of Temporal Power Spectra}\label{exp:temporal_power_spectra}
The Temporal Power Spectra (TPS) of both IPSL-DCPP and ArchesClimate are compared for the same two decades. We use anomalies and use the same variable subset as in \cref{exp:tropical_regional_accuracy}. We include TPS of DiffESM and MESMER-M where possible.

Let us first consider the comparison to MESMER-M and DiffESM. A notably exaggerated monthly signal is evident in DiffESM; its power spectrum exhibits excess power across almost all frequencies, except at the very highest. This discrepancy arises because the model fails to closely follow the seasonal cycle, instead predicting a wide range of values for each month. Consequently, its power spectral density (PSD) aligns more closely with the annual signal, which is consistent with the RMSE and ESR results presented in \cref{sec:generative_rmse_esr}. Conversely, MESMER-M underestimates power slightly, though it does so proportionally across the entire spectrum.

The performance of the generative models becomes particularly clear in this context. We observe that 2D-S768-CL exhibits excess power in \textit{wap}, as well as in the high frequencies of \textit{tos} and \textit{thetaot2000}. In \textit{ta}, 2D-S768-CL overestimates power at high frequencies but shows lower power over longer timescales.

We can directly compare these results to 2D-S768-NCL, the same architecture but trained without the composite loss. This model matches the PSD of the IPSL most closely, confirming that removing these loss terms is a more effective approach to improving variance. Meanwhile, the models using ensembles of deterministic architectures exhibit low power across all variables, with the exceptions of 2D-E192-CL and 2D-E768-CL. While 2D-E192-CL shows high power at high frequencies within the atmospheric variables, increasing the embedding size (2D-E768-CL) eliminates this high-frequency noise. This reinforces our hypothesis that constrained model complexity inhibits the network from properly learning atmospheric dynamics. As in \cref{exp:interchangeability}, we again see strong agreement between 3D-S768-CL and 2D-S768-NCL. We hypothesize that the large embedding size as well as the axial attention allows the model to fix the overpowered features that the composite loss introduces. Based on these findings, the remainder of this analysis will focus exclusively on the 2D-S768-NCL model.

\begin{figure}[H]
    \centering
    \includegraphics[width=\textwidth]{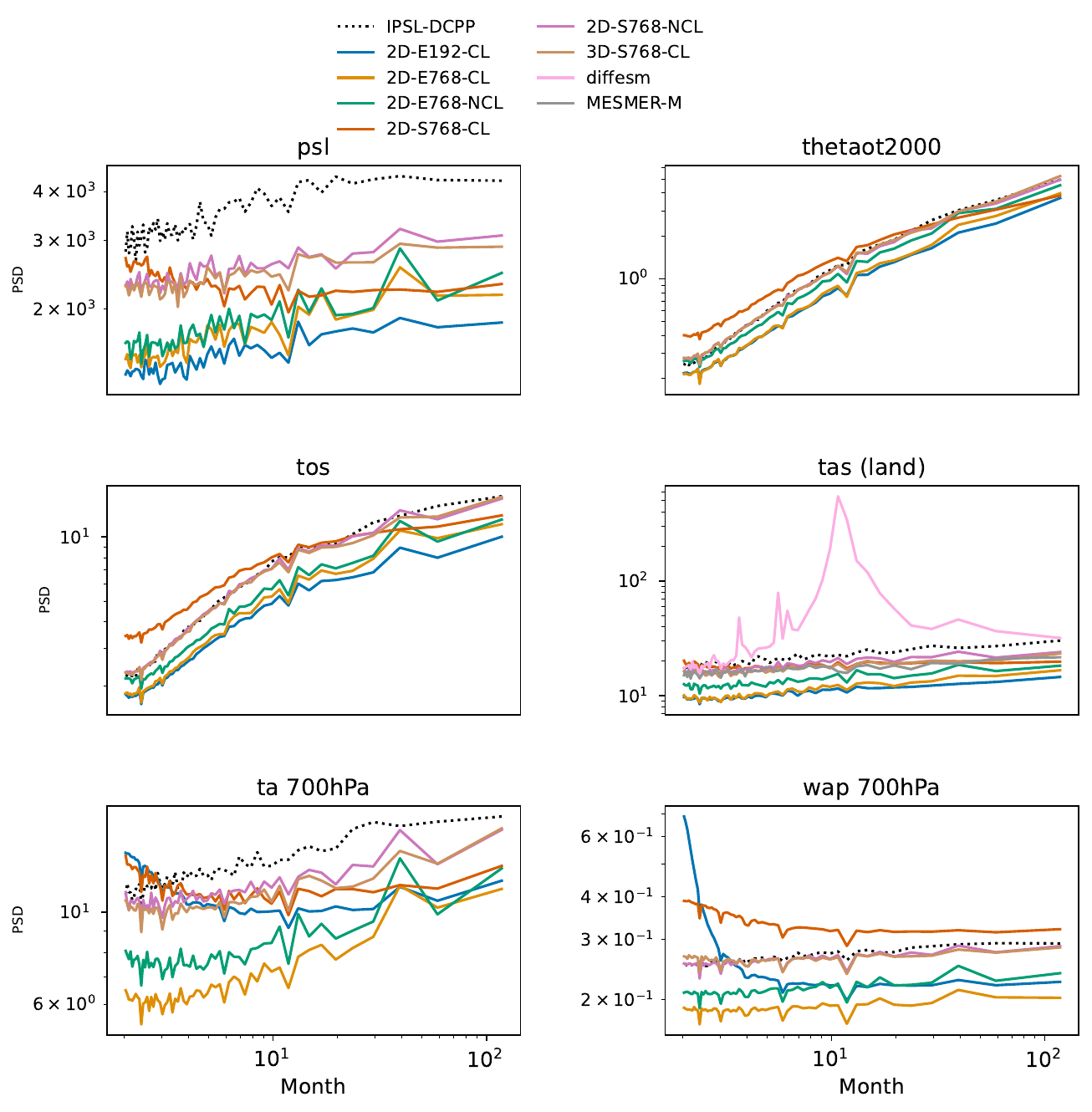}
    \caption{Comparison of the temporal power spectral density of anomalies for the test period 1969--1979 and 2010-2020 across 5-member ensembles of ArchesClimate and IPSL-DCPP. The x-axis is logarithmic, with the smallest value representing one month.}
    \label{fig:temporal_psd}
\end{figure}

\subsection{Regional Accuracy of ArchesClimate in the Tropics}\label{exp:tropical_regional_accuracy}
By training on the full state of a given timestep, ArchesClimate can learn both the seasonal cycle and the deviations from the seasonal cycle. Examining anomalies relative to the seasonal cycle is important to detect trends and patterns of variability. In this experiment, we qualitatively compare the performance of ArchesClimate to IPSL-DCPP in the Tropics (20°S–20°N, 0°–360°E) for both the raw variables and the deseasonalized anomalies. We use ArchesClimate to generate a 5-member ensemble and compare it to a 5-member ensemble of IPSL-DCPP for the test period initialized at 1969 for the Tropics. We target the Tropics as they are influenced by external forcings as well as major modes of internal variability \citep{wangThreeoceanInteractionsClimate2019}. To show the performance in both ocean and atmosphere, we select a subset of variables emulated in ArchesClimate. Sea surface temperature (\textit{tos}) and sea level pressure (\textit{psl}) exemplify atmospheric dynamics, while net heat flux (\textit{net\_flux}, positive downward) shows interactions between atmosphere and ocean. Air temperature (\textit{ta}) at 700 hPa indicates the temperature of the lower free troposphere, which can be influenced by atmospheric convection \cite{hartmannTropicalConvectionEnergy2001}. 

\begin{figure}[H]
    \centering
    % Restrict each image to a maximum of 40% of the page height
    \includegraphics[height=0.45\textheight, keepaspectratio]{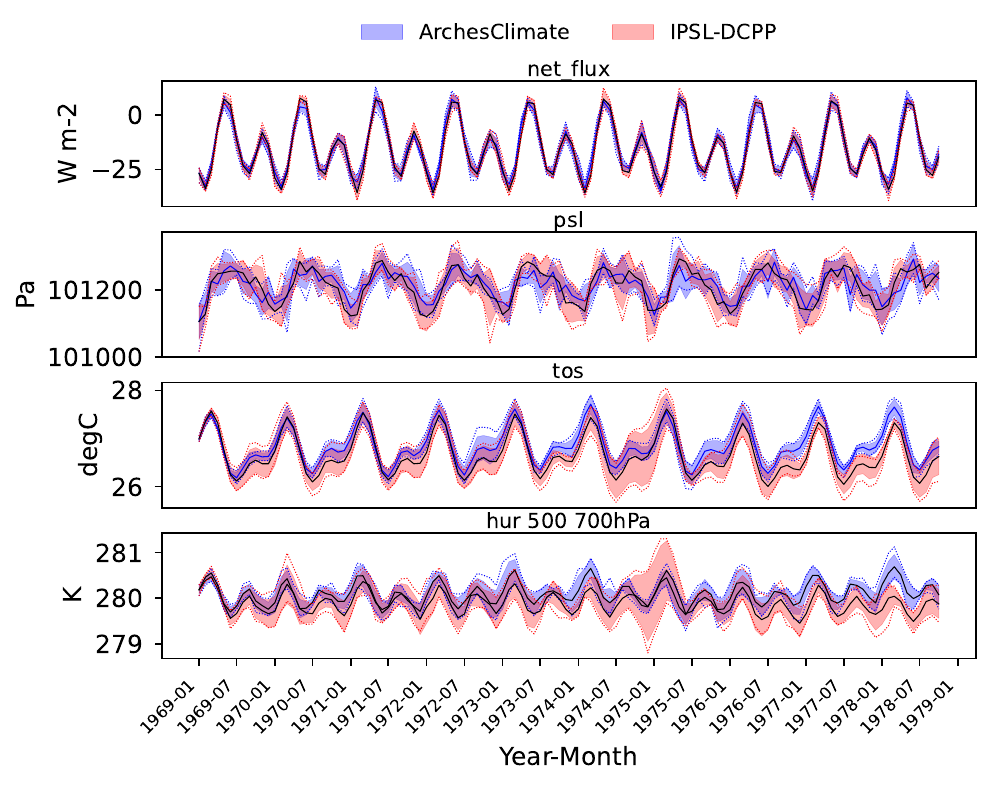}
    
    \vspace{0.5cm} % Adds a small gap between the two images
    
    \includegraphics[height=0.45\textheight, keepaspectratio]{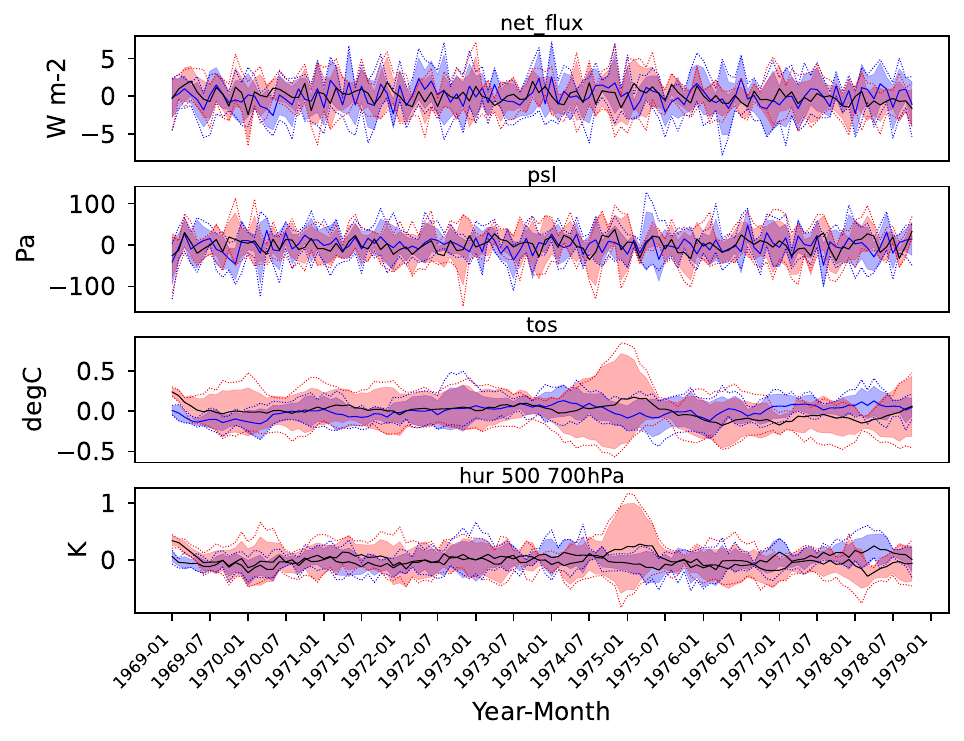}
    
    \caption{Ensemble means for anomalies of the Tropics (20° S – 20° N, 0° – 360° E) for the test period 1969-1979. The dotted lines are the maximum and minimums for each 5-member ensemble mean, with the shaded area being +/- 1 standard deviation. Closer to the dotted line is better.}\label{fig:regional_ensemble_means}
\end{figure}

In \cref{fig:regional_ensemble_means}, ArchesClimate captures the seasonal cycle across the selected variables. When the seasonal cycle is removed, we can see that ArchesClimate produces anomalies of a similar magnitude to IPSL-DCPP. There are small deviations of 0.1-0.2 degrees between ArchesClimate and IPSL-DCPP at the end of the 10 years in the temperature variables \textit{tos} and \textit{ta}, and significant variance around 1975 in IPSL-DCPP that is not present in ArchesClimate. As ArchesClimate is a generative model, internal variability will be present in each ensemble member, providing deviations from IPSL-DCPP.

% \begin{figure}[H]
%     \centering
%     \begin{minipage}[t]{\textwidth}
%         \centering
%         \subfloat{\includegraphics[width=0.9\textwidth]{figures/headline_var_rollout_tropics.pdf}}
%     \end{minipage}
%     \begin{minipage}[t]{\textwidth}
%         \centering
%         \subfloat{\includegraphics[width=0.9\textwidth]{figures/headline_var_rollout_tropics_anomalies.pdf}}
%     \end{minipage}
%    \caption{Ensemble means for raw variables (top) and ensemble means for anomalies (bottom) of the Tropics (20° S – 20° N, 0° – 360° E) for the test period 1969-1979. The dotted lines are the maximum and minimums for each 5-member ensemble mean, with the shaded area being +/- 1 standard deviation. Closer to the dotted line is better.}\label{fig:regional_ensemble_means}
% \end{figure}

\subsection{Long-term Forcing Response}\label{exp:long_term_forcing}

As a final experiment, we emulate 50 years to test the ability of ArchesClimate to respond to external forcings and remain stable over 50 years. We compare this simulation to an identical simulation using external forcings for the year 1969 are repeated every year for 50 years. We take the first ensemble of the validation period (the ensemble initialized at 1969) and the first ensemble of the test period (the ensemble initialized at 2010) to mark the beginning and the end of the 50-year rollout. The 50-year rollout is not presented as a standalone alternative to forced-response GCM runs, but instead we present it as a test of stability. 

\begin{figure}[H]
\includegraphics[width=\textwidth]{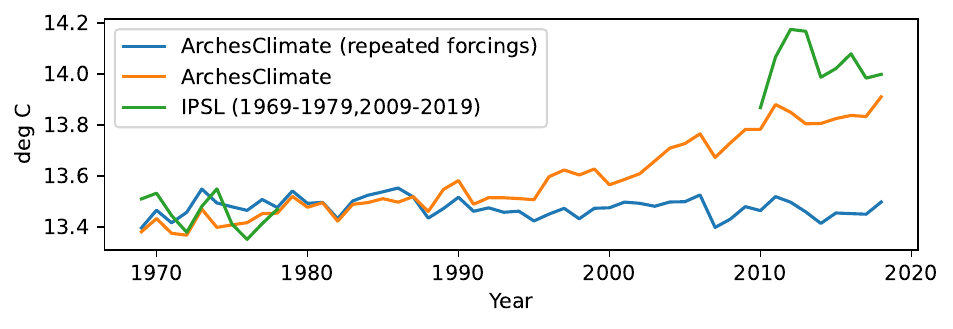}
{\caption{Yearly means for \textit{tos} (sea surface temperature) of a 50-year emulation of ArchesClimate with and without constant external forcings.} 
\label{fig:50_year_rollout}}
\end{figure}

\cref{fig:50_year_rollout} shows the results for sea surface temperature, where ArchesClimate is much closer to the trend of the dataset than the rollout with repeated forcings. Even with limited forcings, there is a noticeable effect on sea surface temperature. Because the forced anthropogenic trend is more detectable at long-terms, this shows that the model responds to subtle boundary conditions without suffering from catastrophic auto-regressive drift or overfitting to internal variability.

\section{Discussion and Conclusions}\label{sec:conclusion}

Our research aims to advance climate modeling by providing a tool to generate ensembles at a reduced cost, enabling more robust probabilistic analysis. We present ArchesClimate and demonstrate its use for the IPSL-CM6A-LR climate model by training on outputs from decadal hindcasts and using CMIP6 greenhouse gas concentrations and solar irradiance as external forcings. We present a combination of a deterministic and a generative machine learning model (based on ArchesWeatherGen) to emulate the climate variability at low computational cost. We also find that the model can auto-regressively produce climatically consistent 10-year sequences at a one-month timestep.

We evaluate ArchesClimate using statistical evaluations and find that it reliably reproduces the mean state and power spectra for several variables and responds well to external forcings. There is a slight bias in several variables, and under-dispersion can be seen in both ESR and rank histograms. We use two methods to address this under-dispersion. The first method, including additional loss terms, introduces too much high frequency noise (as seen in \autoref{fig:temporal_psd}) and the ESR is erroneously inflated due to this. The other method (using a single deterministic model), while still under-dispersive compared to IPSL-DCPP, improves the interchangeability and power spectra of the model. There is improvements needed on variability of the model, as it is an integral part of properly representing the climate. ArchesClimate therefore presents a introduction to coupled atmospheric and oceanic dynamics at a monthly timescale.

In future work, it is possible that ArchesClimate could benefit from a more thorough set of forcings and input variables (e.g. sea ice, land-use change). Similarly, a running average of the last decade's climate could help the model as a sort of memory, especially for variables that operate on longer timescales \citep{mignotDecadalPredictionSkill2016}. By better constraining the model through a memory mechanism, we would expect to produce a more constrained output. 

Our objective was to capture the distribution of each physical variable so that we can generate samples of a given climate resolved at shorter timescales. This task is difficult as it requires expertise in climate science. To aid the assessment of samples, we proposed a set of evaluations of the AI-generated ensembles that can be used in future evaluations of probabilistic climate model emulators. Further work can be done to apply already existing tools such as PCMDI, ESMValTool and climpred \citep{righiEarthSystemModel2020,leeSystematicObjectiveEvaluation2024,bradyClimpredVerificationWeather2021}. With more evaluations, we can increase confidence in the ability of climate model emulators to augment climate models. Analyses can also be done to see if the climate model emulator adheres to conservation properties such as hydrostatic equilibrium and a closed water cycle \citep{shaImprovingAIWeather2025,whiteProjectedNeuralDifferential2024,watt-meyerACE2AccuratelyLearning2025}. 

Despite the encouraging results, this work is subject to limitations due to the use of monthly averages. Primarily, the temporal resolution prevents analysis of daily or sub-daily climate extremes. This modeling choice was motivated by the higher predictability of inter-annual to decadal timescales \citep{boerDecadalClimatePrediction2016}, and the computational gains of predicting climate states at a monthly temporal resolution. By using monthly means, we miss capturing all sub-monthly processes, such as deep convection. We are lacking variance in variables such as \textit{psl} and the model likely underestimates the variance of other variables. 
The use of a monthly temporal resolution may introduce significant biases by smoothing out processes that drive fundamental atmospheric physics. For variables like precipitation ($pr$) and vertical velocity ($wap$), monthly averages mask the discrete nature of convective events, failing to capture the intense latent heat release and moisture transport associated with the convective lifecycle \citep{bonyDynamicThermodynamicComponents2004}. Furthermore, because surface exchange processes are inherently non-linear, calculating variables such as evaporation ($evspsbl$) and ocean-atmosphere net flux ($net\_flux$) from monthly means could overlook submonthly processes that occur within these variables \citep{heldRobustResponsesHydrological2006}. Finally, we cannot estimate the frequency and magnitude of climate extremes, such as heatwaves or frost days, which are critical for assessing ecological and physical impacts.

Furthermore, ArchesClimate directly inherits the biases present in the training climate model data from the IPSL-CM6A-LR. A significant issue is the underprediction of variance across several critical variables, which may limit the model's ability to capture extreme events. Furthermore, the set of state variables contains only a small proportion of the oceanic variables, which limits the interactions between the atmosphere and the ocean that can be studied. We trained a single model on a short time period that contained a large amount of data to better evaluate the model's learning limits. Finally, the ability to conduct comprehensive model ablations and hyperparameter sensitivity analyses was constrained by the high computational resource requirements necessary for training and evaluation. 

Our work opens several new directions of research. Using members generated from ArchesClimate, we believe it will be possible to temporally and spatially downscale these members using similar generative methods to enable analysis at higher resolutions. Further work also includes training on longer experiments such as the Coupled Model Inter-comparison Project experiment \textit{historical} simulations that span several hundred years. By extending ArchesClimate to multi-decadal climate simulations, we can assess if it can emulate inter-annual variability at long timescales \citep{jainImportanceInternalVariability2023}. Expanding the forcings already used in ArchesClimate, interpolation between longer experiments can be investigated to help climate scientists explore previously untested future climate scenarios. The inclusion of other forcings, such as aerosols, which are potentially important for decadal hindcasts \citep{borchertSkillfulDecadalPrediction2021}, is left for further work.

Another possibility is to explore recent advances in generative methods, for which inference time can be done in a fraction of the time, sometimes using only a single inference step \citep{hessFastScaleadaptiveUncertaintyaware2025,schmittConsistencyModelsScalable2024}, which would cut the cost of generating samples by an order of magnitude. Our short investigation into the ability of ArchesClimate to stably generate states for longer temporal periods (see \cref{exp:long_term_forcing}) suggest that emulating climate states at a monthly resolution is an efficient way to predict long-term climate dynamics.  It remains an open question whether ArchesClimate is primarily capturing the underlying processes that operate at shorter timescales, or whether some of its skill at monthly prediction may arise from correlations present at those scales.

ArchesClimate offers an alternative to climate model emulation at a monthly timestep using deep learning. This advancement enables broader access to probabilistic climate projections and supports more timely, informed decision-making in response to climate change.

\appendix

\section{Derivation of \textit{net\_flux} }\label{appendix:netflux}
This section provides the derivation of the variable \textit{net\_flux} as utilized in ArchesClimate. We define $\textit{net\_flux}$ as: 
\begin{equation}
    net\_flux = rsus-rsds+rlus-rlds+hfss+hfls
\end{equation} 
Where $rsus$ represents the Surface Upwelling Shortwave Radiation, $rsds$ is the Surface Downwelling Shortwave Radiation, $rlus$ denotes the Surface Upwelling Longwave Radiation, and $rlds$ signifies the Surface Downwelling Longwave Radiation. The remaining terms, $hfss$ and $hfls$, represent the Surface Upward Sensible Heat Flux and the Surface Upward Latent Heat Flux, respectively.

\section{Alternative Train/Test Splits}\label{exp:alternative_train_test}
\begin{figure}[H]
\includegraphics[width=\textwidth]{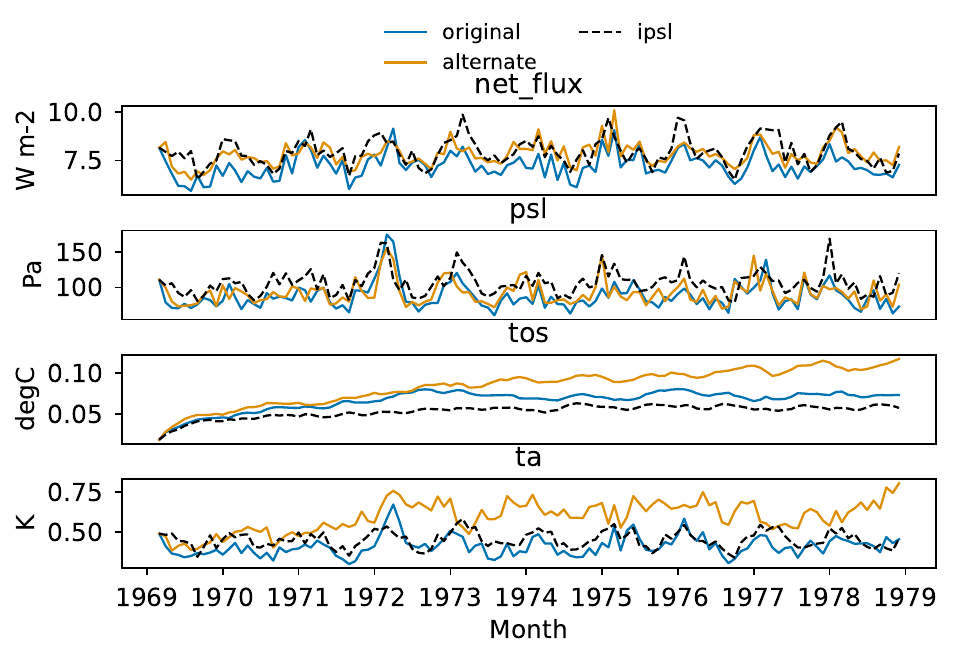}
{\caption{Comparison of CRPS for different train/test splits. ``original" indicates the training scheme outlined in \cref{sec:training} and ``alternative" indicates the training scheme where training and test contain no temporal overlap. Better values are closer to the dotted line, which is the 5-member IPSL-DCPP baseline. We compare 5-member ensembles for the test period 2010-2020. }\label{fig:training_split}}
\end{figure}

In the train/test split of IPSL-DCPP described in \cref{sec:training}, there is temporal overlap in the training and test sets. Here, we look at a train/test split that leaves a test set temporally apart from the train set. To do this, we use the initialization years 1960-2000, so the last year of training is 2009, which we refer to in \cref{fig:training_split} as \textit{alternate}. We can then compare a model that has seen the years 2010-2020 to a model that has not. We generate a 5-member ensemble initialized in 2010 for both models.

We compare CRPS in \cref{fig:training_split}. There is a noticeable difference in CRPS between ArchesClimate and IPSL-DCPP for the variables \textit{tos} and \textit{ta} towards the end of the decade. We speculate that the large influence of external forcings such as the greenhouse gases in 2010-2020 are not well captured when this period is not included in the training. This may deteriorate the generalization of ArchesClimate. ArchesClimate performs much better when it has seen samples from the period of generation. Further investigation is needed to understand if a more comprehensive set of external forcings including the influence of aerosol changes will improve the ability of ArchesClimate to extrapolate to unseen futures. We use the original train/test split for several reasons: our goal in this research is to augment the IPSL-CM6A-LR and not to extend the dataset beyond its current period.  

\clearpage
%%%%%%%%%%%%%%%%%%%%%%%%%%%%%%%%%%%%%%%%%%%%%%%%%%%%%%%%%%%%%%%%%%%%%
% ACKNOWLEDGMENTS
%%%%%%%%%%%%%%%%%%%%%%%%%%%%%%%%%%%%%%%%%%%%%%%%%%%%%%%%%%%%%%%%%%%%%
\acknowledgments
G. Clyne, G. Couairon, and C. Monteleoni were primarily supported by the Choose France Chair in AI, from the French government. The authors thank Juliette Mignot, David Landry, Clément Dauvilliers, and Renu Singh for several discussions about the project. To process the data from IPSL, this study benefited from the IPSL
mesocenter ESPRI facility, which is supported by CNRS, UPMC, Labex L-IPSL, CNES and Ecole Polytechnique. 

%%%%%%%%%%%%%%%%%%%%%%%%%%%%%%%%%%%%%%%%%%%%%%%%%%%%%%%%%%%%%%%%%%%%%
% DATA AVAILABILITY STATEMENT
%%%%%%%%%%%%%%%%%%%%%%%%%%%%%%%%%%%%%%%%%%%%%%%%%%%%%%%%%%%%%%%%%%%%%
% 
%
\datastatement

Code for the project can be found at \href{https://github.com/INRIA/geoarches}{https://github.com/INRIA/geoarches}. The dataset used for training can be found at \href{https://esgf-node.ipsl.upmc.fr/projects/cmip6-ipsl/}{https://esgf-node.ipsl.upmc.fr/projects/cmip6-ipsl/}. 
%  The data availability statement is where authors should describe how the data underlying 
%  the findings within the article can be accessed and reused. Authors should attempt to 
%  provide unrestricted access to all data and materials underlying reported findings. 
%  If data access is restricted, authors must mention this in the statement. See
%  {http://www.ametsoc.org/PubsDataPolicy} for more info.

%%%%%%%%%%%%%%%%%%%%%%%%%%%%%%%%%%%%%%%%%%%%%%%%%%%%%%%%%%%%%%%%%%%%%
% APPENDIXES
%%%%%%%%%%%%%%%%%%%%%%%%%%%%%%%%%%%%%%%%%%%%%%%%%%%%%%%%%%%%%%%%%%%%%
%
%% If only one appendix, use

%\appendix

%% If more than one appendix, use \appendix[<letter>], e.g.,

%\appendix[A] 

%% Appendix title is necessary! For appendix title:

%\appendixtitle{Title of Appendix}

%%% Appendix section numbering (note, skip \section and begin with \subsection)
%
% \subsection{First primary heading}

% \subsubsection{First secondary heading}

% \paragraph{First tertiary heading}

%%%%%%%%%%%%%%%%%%%%%%%%%%%%%%%%%%%%%%%%%%%%%%%%%%%%%%%%%%%%%%%%%%%%%
% REFERENCES
%%%%%%%%%%%%%%%%%%%%%%%%%%%%%%%%%%%%%%%%%%%%%%%%%%%%%%%%%%%%%%%%%%%%%
% Make your BibTeX bibliography by using these commands:
\bibliographystyle{ametsocV6}
\bibliography{archesclimate_bibliography}

\end{document}